\documentclass{aa}
\usepackage{natbib}
\bibpunct{(}{)}{;}{a}{}{,}
\usepackage{pifont}
\newcommand{\xmark}{\ding{55}}
\usepackage{orcidlink}
\usepackage{graphicx}
\usepackage{txfonts} 
\usepackage{hyperref}
\usepackage{caption}
\begin{document}

   \title{Investigating the off-axis GRB afterglow scenario for extragalactic fast X-ray transients}

   \author{H. C. I. Wichern \orcidlink{0009-0004-1442-619X}\inst{1,2}\fnmsep\thanks{hawic@space.dtu.dk}
          \and
          M. E. Ravasio\orcidlink{0000-0003-3193-4714}\inst{1,3}
          \and
          P. G. Jonker\orcidlink{0000-0001-5679-0695}\inst{1,4}
          \and
          J. A. Quirola-Vásquez\orcidlink{0000-0001-8602-4641}\inst{6,7,8}
          \and
          A. J. Levan\orcidlink{0000-0001-7821-9369}\inst{1,5}
          \and
          F. E. Bauer\orcidlink{0000-0002-8686-8737}\inst{6,7,9}
          \and
          D. A. Kann\orcidlink{0000-0003-2902-3583}\inst{10,11}\thanks{Deceased}
          }

   \institute{Department of Astrophysics/IMAPP, Radboud University, PO Box 9010, 6500 GL Nĳmegen, The Netherlands
         \and
             DTU Space, National Space Institute, Technical University of Denmark, Elektrovej 327/328, DK-2800 Kongens Lyngby, Denmark
         \and
             INAF–Astronomical Observatory of Brera, via E. Bianchi 46, I-23807 Merate, Italy
         \and
            SRON Netherlands Institute for Space Research, Niels Bohrweg 4, 2333 CA Leiden, The Netherlands
         \and
            Department of Physics, University of Warwick, Coventry, CV4 7AL, UK
         \and
            Instituto de Astrof{\'{\i}}sica, Pontificia Universidad Cat{\'{o}}lica de Chile, Casilla 306, Santiago 22, Chile
         \and
            Millennium Institute of Astrophysics, Nuncio Monse{\~{n}}or S{\'{o}}tero Sanz 100, Of 104, Providencia, Santiago, Chile
         \and
            Observatorio Astron{\'{o}}mico de Quito, Escuela Polit{\'{e}}cnica Nacional, 170136, Quito, Ecuador
         \and
            Space Science Institute, 4750 Walnut Street, Suite 205, Boulder, Colorado 80301
         \and
            Hessian Research Cluster ELEMENTS, Giersch Science Center, Max-von-Laue-Straße 12, Goethe University Frankfurt, Campus Riedberg, 60438 Frankfurt am Main, Germany
         \and
            Instituto de Astrofísica de Andalucía (IAA-CSIC), Glorieta de la Astronomía, 18008 Granada, Spain
             }

   \date{Received ...; accepted ...}
 
  \abstract
   {Extragalactic fast X-ray transients (FXTs) are short-duration ($\sim$ ks) X-ray flashes of unknown origin, potentially arising from binary neutron star (BNS) mergers, tidal disruption events, or supernova shock breakouts.}
   {In the context of the BNS scenario, we investigate the possible link between FXTs and the afterglows of off-axis merger-induced gamma-ray bursts (GRBs).}
   {By modelling well-sampled broadband afterglows of 13 merger-induced GRBs, we make predictions for their X-ray light curve behaviour had they been observed off-axis, considering both a uniform jet with core angle $\theta_{\rm C}$ and a Gaussian-structured jet whose edge lies at an angle $\theta_{\rm W} = 2\theta_{\rm C}$. We compare their peak X-ray luminosity, duration, and temporal indices $\alpha$ (where $F \propto t^{\alpha}$) with those of the currently known extragalactic FXTs.}
   {Our analysis reveals that a slightly off-axis observing angle of $\theta_{\text{obs}}\approx (2.2-3)\theta_{\rm C}$ and a structured jet are required to explain the shallow ($|\alpha|\lesssim$ 0.3) temporal indices of the FXT light curves, which cannot be reproduced in the uniform-jet case at any viewing angle. 
In the case of a structured jet with truncation angle $\theta_{\rm W} = 2\theta_{\rm C}$, the distributions of the duration of the FXTs are consistent with those of the off-axis afterglows for the same range of observing angles, $\theta_{\text{obs}}\approx (2.2-3)\theta_{\rm C}$. While the distributions of the off-axis peak X-ray luminosity are consistent only for $\theta_{\text{obs}} = 2.2\theta_{\rm C}$, focussing on individual events with different intrinsic luminosities reveals that the match of all three properties (peak X-ray luminosity, duration and temporal indices) of the FXTs at the same viewing angle is possible in the range $\theta_{\text{obs}} \sim (2.2-2.6)\theta_{\rm C}$. Despite the small sample of GRBs analysed, these results show that there is a region of the parameter space -- although quite limited -- where the observational properties of off-axis GRB afterglow can be consistent with those of the newly discovered FXTs.
Future observations of FXTs discovered by the recently launched Einstein Probe mission and GRB population studies combined with more complex afterglow models will shed light on this possible GRB-FXT connection, and eventually unveil the progenitors of some FXTs.}
{}

   \maketitle

\section{Introduction}

Over the last decade, a new class of X-ray transients has emerged from \textit{Chandra}, \textit{XMM-Newton}, {\em Swift} and \textit{eROSITA} archival observations in the soft X-ray regime \citep{2008Soderberg, 2013Jonker, 2015Glennie, 2016Irwin, 2017Bauer, 2019Xue,  2020Wilms, 2020AlpLarsson, 2020Novara, 2022Lin, 2022QuirolaVasquez, 2023QuirolaVasquez}. These fast X-ray transients (FXTs) are characterised by short durations of a few minutes to hours. Given the broad variety in their light curve behaviour, spectra, and host galaxies, it is considered likely that they originate from a varied set of progenitors rather than a single class. Determining their origins is not trivial, however, since they lack counterparts at other wavelengths and robust redshift estimates in most cases, despite thorough searches for potential host galaxies \citep{2017Bauer, 2019Xue, 2019Lin, 2021Lin, 2021Andreoni, 2021Jonker, 2022Lin, 2022Eappachen, 2022QuirolaVasquez, 2023Eappachen, 2023QuirolaVasquez}. Possible origins include supernova shock breakouts, tidal disruptions of a white dwarf by an intermediate-mass black hole, or mergers of two neutron stars (e.g. \citealt{2020AlpLarsson, 2013Jonker, 2017Bauer, 2019Xue, 2019DadoDar, 2020DadoDar}), of which the latter may be accompanied by a gamma-ray burst (GRB) and its broadband afterglow (e.g. \citealt{1989Eichler, 1992Narayan, Nakar2007, 2017Abbott, 2020Gompertz}).  
\\
Seventeen distant ($>$ 100 Mpc) FXTs were (re-)discovered in \textit{Chandra} archival data (2000 -- 2022) by \citet{2022QuirolaVasquez} and \citet{2023QuirolaVasquez}. A few of them share light curve, spectral, and potential host galaxy properties with those observed for X-ray afterglows of merger-induced GRBs (GRBs of compact-object merger origin). However, these FXTs (1) are less luminous than typical GRB afterglows (with a peak X-ray luminosity of $\sim 10^{44-47}$ erg/s), (2) lack associated prompt $\gamma$-ray emission, and (3) often exhibit a rising phase or flat segment (plateau) in their light curves, which is not expected in the standard afterglows of GRBs if they are assumed to be observed on-axis. This may suggest that some FXTs could be the X-ray afterglows of merger-induced GRBs that are observed off-axis (such that the line of sight lies outside the core of the relativistic jet), especially if these GRBs possess a structured jet rather than a conventional uniform jet. Indeed, a structured jet has been suggested previously to produce a plateau in the light curves of GRB afterglows observed off-axis (e.g. \citealt{2020Oganesyan, 2020Ascenzi, 2020Beniamini, 2020Ryan, 2021TakahashiIoka}). An early attempt to explore this possible connection has been reported in \citealt{2021Sarin}, where the X-ray data of one FXT (CDF-S XT1; \citealt{2017Bauer}), despite their challenging signal-to-noise, have been fitted with an afterglow produced by a structured jet viewed off-axis, and also in \citealt{2023QuirolaVasquez}, where by comparing synthetic afterglow light curves to those of FXTs they concluded that a slightly off-axis short GRB scenario is plausible.
\\ 
In this work, we investigate the possible connection between extragalactic FXTs and the afterglows of off-axis merger-induced GRBs. We consider a sample of 13 merger-induced GRBs whose afterglow emission has been well-sampled through multi-wavelength observations and obtained their afterglow parameters by fitting them with the semi-analytical model \texttt{afterglowpy} \citep{2020Ryan}. With the information obtained from the fitted on-axis afterglow, we build the expected light curves for the off-axis case and derive the related properties (in particular the luminosity, duration, and temporal indices) as a function of the viewing angle, considering both a uniform- and a structured-jet model. We then compare the values obtained from the off-axis merger-induced GRB afterglows at different viewing angles with the measured FXT parameters. 
In light of the lack of optical counterparts to FXTs, we also compare the available optical upper limits to the predicted off-axis afterglows in the same band at different viewing angles. In Sects.~\ref{sec:Selection} and~\ref{sec:Methods} we describe our FXT and merger-induced GRB sample selection and our methods, in Sect.~\ref{sec:Results} we describe our results, and in Sect.~\ref{sec:Discussion} and~\ref{sec:Conclusion} we provide a discussion of the results and summarise the main findings, respectively. Throughout this paper, we assume a flat $\Lambda CDM$ cosmology with $H_{0}$ = 67.4 $\pm$ 0.5 km s$^{-1}$ Mpc$^{-1}$ and $\Omega_{m}$ = 0.315 $\pm$ 0.007 \citep{2018Planck}.

\section{Sample selection}
\label{sec:Selection}
\subsection{FXTs}
We based our selection of the FXT sample on the works of \citet{2022QuirolaVasquez, 2023QuirolaVasquez}, which provide the most updated and comprehensive catalogue of FXTs discovered so far in \textit{Chandra} data.
Of the 22 FXTs reported by \citet{2022QuirolaVasquez, 2023QuirolaVasquez}, five FXTs belong to a \textit{nearby} sample located within a distance of 100 Mpc. We focus only on the sample of 17 \textit{distant} FXTs, of which seven have redshift estimates (which lie in a range $z$ = 0.61--2.23). For the 10 FXTs without a redshift measurement, the same fiducial redshift as adopted by \citet{2022QuirolaVasquez, 2023QuirolaVasquez} is assumed (which lie in a range $z$ = 0.0216--1.0; see Table~\ref{tab:FXRT} in Appendix~\ref{sec:appendix-GRBs}).

\subsection{Merger-induced GRBs}
The GRBs analysed in this work have been selected based on the sampling of their multi-wavelength afterglow light curve. Historically, short-duration GRBs ($T_{90}<$ 2 s \footnote{$T_{90}$ is defined as the time interval over which the cumulative number of counts above the background increases from 5\% to 95\% \citep{1993Kouveliotou}.}) have been attributed to compact-object mergers (e.g. \citealt{1989Eichler,Nakar2007} and references therein). However, there is now compelling evidence that some GRBs with $T_{90} >$ 2\ s are merger-driven (e.g. \citealt{2022Rastinejad, 2022Troja, 2023Gompertz, 2023Levan}) rather than of a collapsar origin. For this reason, instead of only considering short-duration GRBs, we use a more lenient classification scheme and focus on GRBs that have been suggested to be of merger origin based on other properties rather than just their duration. 
We started with an initial sample consisting of 48 GRBs for which optical data were available, collected by D.A. Kann from literature (see Table~\ref{tab:sGRB} in Appendix~\ref{sec:appendix-GRBs} for all literature references). The first 37 of these 48 GRBs correspond to the sample of Type I GRBs reported in Table 2 of \citet{2011Kann}\footnote{Except for GRB 071112B and the anomalous burst GRB 060121, which are not part of our initial sample.}, who used the classification scheme by \citet{2009Zhang} to classify Type I (compact-object merger origin) and Type II (collapsar origin) GRBs. The flowchart by \citet{2009Zhang} classifies Type I and Type II GRBs based on their (rest-frame) duration $T_{90}$, whether there is a supernova association or not, the properties of their host galaxies (the host being early- or late-type, the specific star formation rate, for instance), the offset from the host, the density and density profile of the surrounding medium, whether they follow the Amati \citep{2002Amati} and Norris \citep{2000Norris} relations, and based on their prompt and kinetic energy $E_{\gamma}$, $E_{K}$, respectively. To this published selection of well-sampled optical merger-induced GRB afterglows, another 11 bursts were added later on by Dr. Kann, of which nine have $T_{90}<2$~s (GRB 100702A, GRB 101219A, GRB 110112A, GRB 111020A, GRB 111117A, GRB 130603B, GRB 160821B, GRB 170817A, and GRB 181123B; \citealt{2012Margutti}, \citealt{2012Fong}, \citealt{2013Fong}, \citealt{2013Cucchiara}, \citealt{2018Zhang}, \citealt{2019Lamb}, \citealt{2020Paterson}) and two have $T_{90}>2$ s and are characterised by an initial short-hard spike followed by a softer extended emission (GRB 150524A and GRB 160410A; \citealt{2022Fong}, \citealt{2023AguiFernandez}). The optical light curves consist of multi-colour observations that were converted to the $R_{c}$ band (see e.g. \citealt{2007Kann,2011Kann}), and the given magnitudes are AB magnitudes which were corrected for Galactic extinction and, where possible, the contributions from the host galaxy. In the case of GRB 160821B, the contribution from a potential kilonova has also been subtracted from the optical light curve by extrapolating the X-ray light curve to the optical band (e.g. \citealt{2019Lamb}). We converted the AB magnitudes to flux densities\footnote{This conversion was done using the \texttt{astropy} \citep{astropy:2013, astropy:2018, astropy:2022} units package, and a central frequency of 6470~\text{\AA} was assigned to the $R_{c}$ band.}. The optical data are reported with 1$\sigma$ uncertainties in the case of detections or as 3$\sigma$ upper limits in the case of non-detections. 
For each of the 48 GRBs, we then searched for X-ray data from the UKSSDC \textit{Swift}/XRT light curve Repository \citep{2007Evans, 2009Evans}. We downloaded the unabsorbed 0.3--10 keV flux light curves (using default binning parameters) of the GRBs detected by \textit{Swift} from the repository. Where available, we collected additional X-ray data as well as radio data from the literature. In total, we obtained radio upper limits for seven bursts and radio detections for two bursts in our final sample of 13 GRBs, for which we will now describe the selection criteria.

Afterglows of GRBs may, at any wavelength, be contaminated by emission components that do not originate only from the forward shock. Examples include the contribution of a reverse shock, a kilonova, or an episode of energy injection which may give rise to plateaus and flares. To model only the forward shock emission of broadband merger-induced GRB afterglows, we had to exclude data points possibly contaminated by non-afterglow emission components and treat them as upper limits (as was also done in e.g. \citealt{2009Perley, 2012Metzger, 2019Troja}). 
In identifying these features, we are aided by the available literature (see Appendix~\ref{sec:appendix-GRBs}) and also by the evolution of the spectral photon index (e.g. Fig.~\ref{fig:fits}), which is indicative of a possible tail of prompt emission origin \citep{2007Zhang}. 
We define a plateau as a light curve segment that has a slope of $|\alpha| <$ 0.8 (e.g. \citealt{2019Zhao, 2022Ronchini}).
After identifying such non-afterglow components, we require the GRBs in our sample to have a total (so considering all bands: X-ray, optical and radio) number of detections greater than nine, in order to have at least two degrees of freedom in the fit (given that the applied afterglow model has seven free parameters). This additional criterion led to our final sample of 13 GRBs, which are reported in Table~\ref{tab:sGRB}. We note that while GRB 170817A did pass all aforementioned criteria, it was left out since it was observed off-axis (e.g. \citealt{2017Abbott}), in order to be consistent in our methods (since we fit the GRB afterglows in our final sample under the assumption that they are observed on-axis). We include the redshift and X-ray spectral photon index measurements of the GRBs in the final sample. The redshifts of the GRBs in our final sample lie in a range $z$ = 0.16--2.211. The time-averaged value of the photon index ($\Gamma_{X}$) was used to convert the unabsorbed 0.3--10 keV fluxes to flux densities at 10 keV. This is done because at this energy, it is possible to verify our converted flux densities with the 10 keV flux densities reported on the webpage of the \textit{Swift}/XRT light curve Repository. These light curves, however, are more sensitive to fluctuations in the photon index, while our light curves are built assuming a constant, average photon index.  The data selection of the final sample is described in more detail in Appendix~\ref{sec:appendix-GRBs}. We note that seven out of 13 GRBs in our final sample have durations $T_{90}$ exceeding 2 s (see Table~\ref{tab:sGRB}).

\section{Methods}
\label{sec:Methods}
\subsection{Broadband afterglow modelling}
For each of the 13 GRBs in our sample, we fit the corresponding multi-wavelength afterglow data with the semi-analytical model \texttt{afterglowpy} v. 7.3.0. \citep{2020Ryan} through a Markov Chain Monte Carlo (MCMC) approach using the
Python package \texttt{emcee} \citep{2013ForemanMackey}. The open-source Python module \texttt{afterglowpy} allows for the calculation of GRB afterglow light curves and spectra, while taking into account both viewing-angle effects and different jet structures. The latter determines how the energy of the jet depends on the polar angle $\theta$. For the purpose of this work, we selected two jet profiles: a uniform jet, where the energy profile is described as
\begin{equation}
    E(\theta) = 
    \begin{cases}
        E_{0} & \theta \leq \theta_{\rm C} \\
        0 & \theta > \theta_{\rm C},
    \end{cases}
\end{equation}
and a structured jet with a Gaussian energy profile, given by
\begin{equation}
    E(\theta) = 
    \begin{cases}
    E_{0}\text{ exp}\left(-\frac{\theta^{2}}{2\theta_{\rm C}^{2}}\right) & \theta \leq \theta_{\rm W} \\
    0 & \theta > \theta_{\rm W}.
    \end{cases}
\end{equation}
For the Gaussian-structured jet model, the set of free parameters is
$\Theta = \{\theta_{\text{obs}}, E_{0}, \theta_{\rm C}, \theta_{\rm W}, n_{0}, p, \epsilon_{e}, \epsilon_{B}, \xi_{N}, d_{L}\}$, where $\theta_{\text{obs}}$ is the viewing angle, $E_{0}$ the isotropic-equivalent kinetic energy of the blastwave, $\theta_{\rm C}$ the core angle of the jet, $\theta_{\rm W}$ the truncation angle that marks the outer edge of the structured wings of the jet, $n_{0}$ the density of the uniform circum-burst medium, $p$ the power-law index of the electron energy distribution, $\epsilon_{e}$ and $\epsilon_{B}$ the fraction of thermal energy that is imparted to electrons and magnetic fields, respectively, $\xi_{N}$ the fraction of electrons that is accelerated at the shock front, and $d_{L}$ the luminosity distance. In the afterglow modelling, the redshift of each burst is kept fixed (see Table~\ref{tab:sGRB}), and the luminosity distance is computed from this redshift. To reduce the number of degrees of freedom, we applied the typical approach of fixing $\xi_{N}$ = 1, often assumed in afterglow modelling due to a degeneracy of $E_{0}$, $n_{0}$, $\epsilon_{e}$, and $\epsilon_{B}$ with $\xi_{N}$ \citep{2005Eichler}. The uniform-jet model has one free parameter fewer than the Gaussian-structured one (seven versus eight free parameters), due to the absence of the truncation angle $\theta_{\rm W}$ parameter. 

When we modelled the afterglows of the GRBs in our sample with a structured-jet model leaving the truncation angle $\theta_{\rm W}$ free in a reasonable range (namely $0 < \theta_{\rm W}\leq$ max($4\theta_{\rm C}, \pi/2$)) and with a uniform-jet model, we found that the afterglow data do not allow us to distinguish between the two models (especially considering the negligible differences between their energy profiles when observed on-axis, when $\theta_{\text{obs}} < \theta_{\rm C}$). We found that $\theta_{\rm C}$ and $\theta_{\rm W}$ were nearly unconstrained and / or had largely overlapping posterior distributions. 
For this reason, for the structured-jet model, we adopt the same afterglow parameters obtained from the fit with the uniform-jet model, and fix the truncation angle at $\theta_{\rm W} = 2\theta_{\rm C}$. This choice is based on a study by \citet{2018BeniaminiNakar} who find that $\gamma$-ray emission is unlikely to be observed outside of a narrow region around the core of the jet ($\theta_{\text{obs}}\gtrsim 2\theta_{\rm C}$ for a Gaussian-structured jet with $\theta_{\rm W}\approx$ 3.7$\theta_{\rm C}$), meaning that energetic wings with a wide truncation angle $\theta_{\rm W} \gg \theta_{\rm C}$ are disfavoured. However, we have repeated the same analysis considering a different value of $\theta_{\rm W} = 1.5\theta_{\rm C}$ to probe different extents of the jet structure. Although this case yields different quantitative results (mainly regarding the luminosity) compared to the $\theta_{\rm W} = 2\theta_{\rm C}$ case, we draw the same qualitative conclusions regarding the comparison between FXTs and off-axis merger-induced GRB afterglows in both cases. Thus, we focus on the results for the $\theta_{\rm W} = 2\theta_{\rm C}$ case in the main text and refer to Appendix~\ref{sec:appendix-alt} for the results of the $\theta_{\rm W} = 1.5\theta_{\rm C}$ case. Table~\ref{tab:prior} shows the \texttt{afterglowpy} model parameters and the form and bounds of their prior distributions assumed for the MCMC sampling (in this Table, we refer to the $\theta_{\rm W} = 2\theta_{\rm C}$ and $\theta_{\rm W} = 1.5\theta_{\rm C}$ cases as Case I and Case II, respectively, as they are referred to in Appendix~\ref{sec:appendix-alt}). We used 64 walkers, and one trial run was used to initialise the walkers and to estimate the number of steps required for each Markov chain to converge. The Geweke statistic \citep{1992Geweke} was used as a convergence diagnostic. A burn-in period with a length of twice the maximum auto-correlation time \citep{2013ForemanMackey} was discarded. We discuss the fit results of the GRB afterglow modelling in Appendix~\ref{sec:appendix-fitting}. 
\begin{table}
\caption{Prior distributions and bounds of model parameters used in the MCMC parameter estimation for the Gaussian-structured jet model.}
\centering
\begin{tabular}{cccc} \hline
    Parameter &  Unit & Prior & Bounds \\ \hline
    $\theta_{\text{obs}}$ & radians & uniform & [0,$\theta_{\rm C}$] \\
    $E_{0}$ & erg & log-uniform & [49, 54] \\
    $\theta_{\rm C}$ & radians & uniform & (0, $\pi$/6] \\
    $n_{0}$ & cm$^{-3}$ & log-uniform & [-6,1]  \\
    $p$ & - & uniform & (2, 3]  \\
    $\epsilon_{e}$ & - & log-uniform & [-4, log(1/3)) \\
    $\epsilon_{B}$ & - & log-uniform & [-8, log(1/3)) \\
    \hline
    $\theta_{\rm W}$ & radians & fixed & 2$\theta_{\rm C}$ (Case I) \\
    $\theta_{\rm W}$ & radians & fixed & 1.5$\theta_{\rm C}$ (Case II) \\
\end{tabular}
\tablefoot{For the uniform-jet model, the prior distributions and bounds are the same as in the Gaussian-structured jet model; only the truncation angle $\theta_{\rm W}$ is missing as a model parameter in the case of a uniform jet. Square brackets indicate the bounds are included in the given interval, while round brackets indicate the bounds are excluded.}
\label{tab:prior}
\end{table}

\subsection{Generating off-axis GRB afterglow light curves}
The best-fit flux density light curves for the on-axis case are transformed to their off-axis counterparts by varying the viewing angle $\theta_{\text{obs}}$ whilst keeping all other jet parameters (obtained via MCMC sampling) fixed. When generating the off-axis light curves, we restrict the posterior distributions of the jet parameters $E_{0}$, $\theta_{\rm C}$, $n_{0}$, $p$, $\epsilon_{e}$, and $\epsilon_{B}$ to the 16th -- 84th percentile range. For a range of different ratios $\theta_{\text{obs}}/\theta_{\rm C}$, we then compute the peak X-ray luminosity in the 0.3--10 keV band and the duration. The duration $t$ of the GRB afterglows is defined as the time interval during which the off-axis afterglow flux is above the assumed flux limit. We set this flux limit to be 10 times the nominal flux limit of the \textit{Chandra}/ACIS instrument, with which the FXTs were detected. This is motivated by the fact that this flux limit allows for a fair comparison with FXTs that were reported in \citet{2022QuirolaVasquez, 2023QuirolaVasquez}, who also assumed a detection threshold ten times the nominal flux limit to avoid source detections due to Poisson fluctuations. However, we repeated the analysis assuming the nominal \textit{Chandra} flux limit and we report the results in the discussion section (see Sec.~\ref{sec:Discussion}).
Adopting a fiducial photon index of $\Gamma_{X}\approx$ 2 and an integration time of $\tau$ = 20 ks (as most FXTs have durations $T_{90}$ of 20 ks or less), we transform the point-source sensitivity of \textit{Chandra}/ACIS multiplied by a factor ten in the 0.4--6.0 keV band\footnote{\url{https://cxc.harvard.edu/proposer/POG/html/chap6.html\#tab:acis_char}} to the 0.3--10 keV band assuming $F_{\text{lim}}\propto \tau^{-1/2}$ (which is valid in case the sensitivity is background-limited). Subsequently, the integrated 0.3--10 keV fluxes are converted to the isotropic rest-frame X-ray luminosity in the same band.
We repeated the same analysis assuming a different integration time of $\tau$ = 10 ks, whose results are in line with our main analysis' results and reported in Appendix~\ref{sec:appendix-alt}.

\begin{figure*}
	\includegraphics[width=\columnwidth]{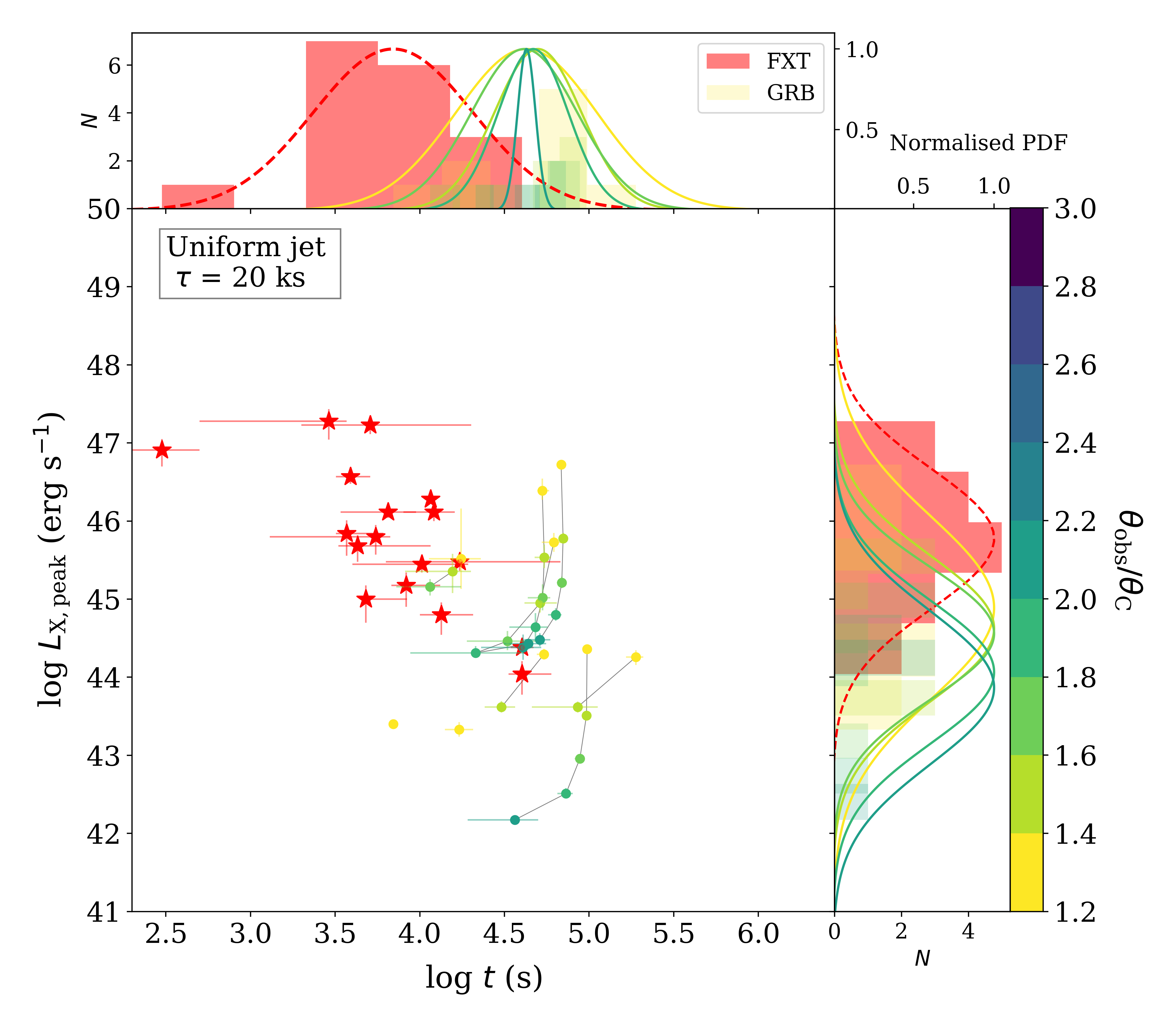}
	\includegraphics[width=\columnwidth]{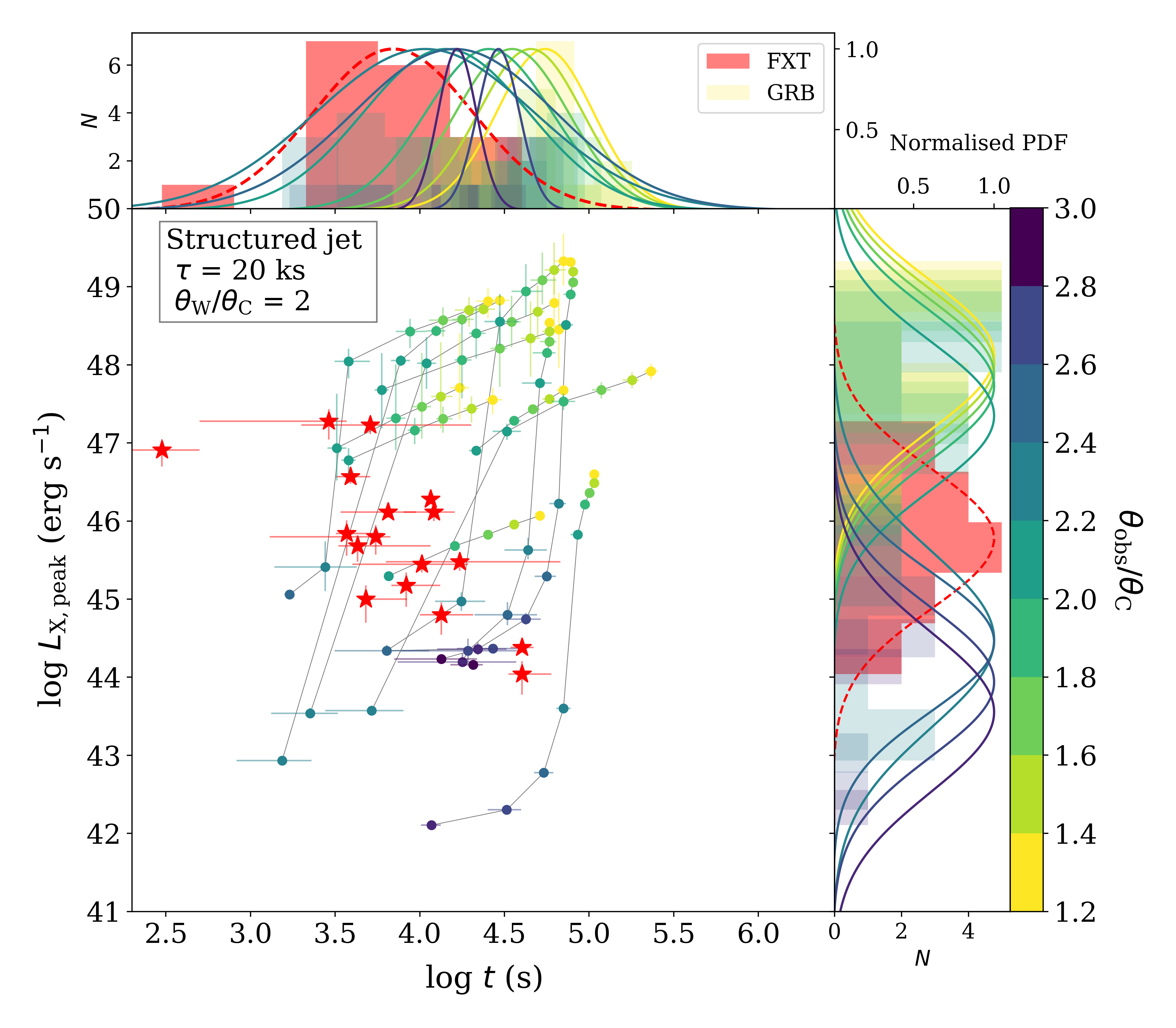}
\caption{Peak X-ray luminosities $L_{X,\text{peak}}$ in the 0.3--10 keV band, and durations $T_{90}$ (Table ~\ref{tab:FXRT}) of distant FXTs (red stars) and durations $t$ above the assumed flux limit of off-axis merger-induced GRB X-ray afterglows (coloured points). The latter is shown for different viewing angles $\theta_{\text{obs}}/\theta_{\rm C}$. The histograms show the sample distributions of the luminosity and duration of FXTs on one hand and GRBs on the other hand; over-plotted on the histograms are the corresponding normalised probability densities (red dashed lines and coloured solid lines corresponding to the sample of FXTs and GRBs, respectively) showing how the distributions of the peak X-ray luminosity and duration shift with increasing viewing angle ($\theta_{\text{obs}}/\theta_{\rm C}$). Results on the left correspond to the uniform-jet model, and those on the right to the structured-jet model with $\theta_{\rm W}$ = 2$\theta_{\rm C}$.}
\label{fig:X-rays}
\end{figure*}

\section{Results}
\label{sec:Results}

\subsection{X-ray luminosity and duration}
\label{sec:X-rays}
Figure ~\ref{fig:X-rays} shows the peak X-ray luminosity and duration of the sample of FXTs and of the sample of synthetic merger-induced GRB afterglows at different viewing angles. We calculated the peak X-ray luminosity and duration for a range of viewing angles given by $\theta_{\text{obs}}/\theta_{\rm C} \in$ [1.2, 1.4, 1.6, 1.8, 2, 3, 4, 5, 6, 7, 8] for both uniform-jet and structured-jet cases, although for the latter we included also the values $\theta_{\text{obs}}/\theta_{\rm C} \in$ [2.2, 2.4, 2.6, 2.8], such that we perform a finer sampling in the off-axis range $\theta_{\text{obs}} = (2-3)\theta_{\rm C}$.

The distributions of the peak X-ray luminosity and duration are shown in the histograms on the top and on the right panels of each plot. We note that although the viewing angle was varied up to a value $\theta_{\text{obs}}= 8\theta_{\rm C}$, no off-axis afterglows remain detectable above our assumed flux limit for $\theta_{\text{obs}}\geq 3\theta_{\rm C}$ even when considering a structured-jet model, such that the colour bar only shows values up to $\theta_{\text{obs}}/\theta_{\rm C}=3$. The left panel shows the results related to the uniform-jet case, while the right panel shows the case of the structured jet with a truncation angle $\theta_{\rm W} = 2\theta_{\rm C}$.

In the uniform-jet case, the left panel shows that the off-axis afterglow light curves of the 13 GRBs in our sample rapidly drop below the flux limit (or equivalently, the luminosity limit) as we go even slightly off-axis, for example at a viewing angle of $\theta_{\text{obs}} = 1.2\theta_{\rm C}$ only nine out of 13 GRBs remain detectable above this limit whereas at $\theta_{\text{obs}} = 2\theta_{\rm C}$ only four out of 13 GRBs remain detectable. Based on the distributions shown on top and to the right of the panel, it appears that the luminosity distribution of the analysed GRB afterglows at for example $\theta_{\text{obs}} = 1.2\theta_{\rm C}$ may be similar to that of the FXTs, while the distributions of their durations are not consistent. 
To assess this quantitatively, we perform a non-parametric Kolmogorov--Smirnov (KS) test between the cumulative distributions of the peak X-ray luminosity and duration of FXTs with those of the off-axis afterglows of the 13 analysed GRBs.

In the uniform-jet scenario, the luminosity distributions of the FXTs and GRB off-axis afterglows have $p$-values of $p = 0.06$ and $p = 0.09$ at $\theta_{\text{obs}} = 1.2, 1.4\theta_{\rm C}$, respectively. Beyond $\theta_{\text{obs}} \geq 1.6\theta_{\rm C}$, the afterglows become too faint compared to the X-ray luminosities of the FXTs ($p\lesssim 0.02$). Comparing their durations yield low probabilities for any viewing angle ($p\lesssim$ 0.03), with the afterglows lasting on average five times longer than FXTs.

The results for the case of a structured jet with a truncation angle fixed at $\theta_{\rm W} = 2\theta_{\rm C}$ are shown in the right panel of Fig. ~\ref{fig:X-rays}. Before we describe the results, we note that four out of 13 GRBs (GRB 051227, GRB 061006, GRB 070724A, and GRB 071227) have particularly large uncertainties on their peak X-ray luminosity at viewing angles $\theta_{\text{obs}}\leq\theta_{\rm W}$. This is because these GRBs have poorly constrained fitted parameters, in particular the isotropic-equivalent kinetic energies $E_{0}$ and circum-burst densities $n_{0}$, which heavily affects the peak luminosity of the afterglow light curve (see \citealt{1999Sari, 2004ZhangMeszaros, 2020Oconnor}).
Compared to the uniform-jet case, the afterglows in the structured-jet case can reach higher luminosities (a difference of 3.4 dex on average) at the same viewing angle when the line of sight lies within the wings of the jet, that is, when $\theta_{\text{obs}}\leq\theta_{\rm W}$. This higher luminosity makes the structured-jet GRB afterglows too bright to be compatible with that of FXTs, and a KS test between the distribution of the peak X-ray luminosities of the sample of FXTs and of the sample of GRB afterglows returns  $10^{-6} \lesssim p\lesssim~10^{-3}$ when $\theta_{\text{obs}}\leq\theta_{\rm W}$. However, as soon as we go slightly outside the wings, there is a single viewing angle $\theta_{\text{obs}} = 2.2\theta_{\rm C}$ for which the distribution of peak luminosities of our GRB off-axis afterglows is compatible with those of the FXTs, at a significance level of $p = 0.10$ (with eight out of 13 GRBs being detected at this viewing angle. Their redshifts lie in a range $z$ = 0.111 -- 2.211). For viewing angles $\theta_{\text{obs}} > 2.2\theta_{\rm C}$, the distributions of luminosities of the off-axis GRB afterglows are again incompatible with those of the FXTs, likely due to the former starting to become too faint.
Regarding the durations, the situation is more promising. While the distributions of the durations of GRB off-axis afterglows are not consistent with those of FXTs when observed within the truncation angle ($\theta_{\text{obs}} < 2\theta_{\rm C}$), as soon as the viewing angle lies outside the wings the duration distributions start to be compatible, as confirmed by a KS test, reporting $p$ = 0.27, 0.33, 0.24 for $\theta_{\text{obs}} = 2, 2.2, 2.4 \, \theta_{\rm C}$, respectively. Although a possible match ($p \sim 0.10 - 0.12$) is also found for $\theta_{\text{obs}} > 2.6\theta_{\rm C}$, at such greater viewing angles the number of GRBs afterglow detectable above the flux limit is progressively lower (four to two, out of 13), hindering a proper comparison. Moreover, as previously mentioned, at such large viewing angles the peak X-ray luminosities of the FXTs and the off-axis afterglows are less likely to be consistent with one another ($p\lesssim 0.04$). 

Summarizing, while in the uniform-jet case both the luminosities and the durations of the GRB off-axis afterglow analysed are not compatible with those of the FXTs, in the structured-jet case with a truncation angle $\theta_{\rm W}$ = 2$\theta_{\rm C}$ there are more overlaps and in particular, the distribution of the durations are consistent for $\theta_{\text{obs}} = (2-2.4)\theta_{\rm C}$ and we find one viewing angle ($\theta_{\text{obs}} = 2.2\theta_{\rm C}$) at which both the peak X-ray luminosity and duration of the two populations are consistent.

\subsection{Temporal indices}
\begin{figure*}
	\includegraphics[width=2\columnwidth]{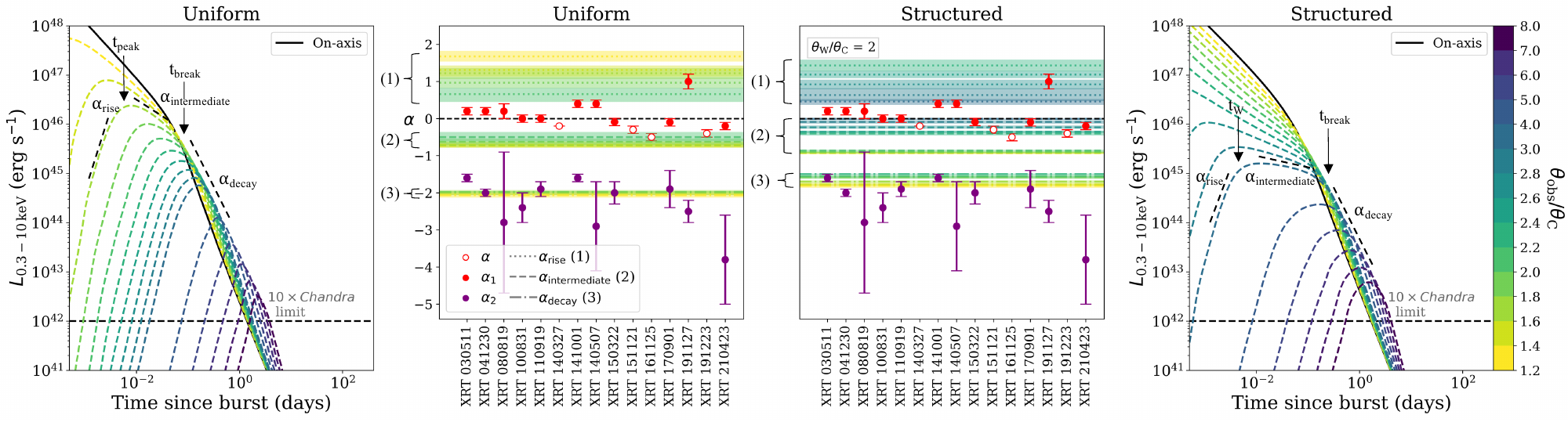}
\caption{Comparison of the mean temporal indices of the off-axis afterglow light curves of the analysed GRB sample observed at different viewing angles (dotted, dashed, and dot-dashed lines) with the observed temporal indices of each FXT (red and purple points). The first and fourth panels show example light curves at different viewing angles for a uniform and a structured jet with $\theta_{\rm W}$ = 2$\theta_{\rm C}$, respectively, and the \textit{Chandra} flux limit (converted to a luminosity limit in this case, assuming a photon index of $\Gamma\approx 2$ and a redshift of $z = 0.2851$) is drawn arbitrarily at $L_{\text{0.3--10 keV}} = 10^{42}$ erg s$^{-1}$, for illustrative purposes.
The results are shown for the uniform-jet case (second panel) and the structured-jet case (third panel). The horizontal lines (plus 16th -- 84th percentile uncertainty regions) correspond to the mean slopes of the off-axis GRB afterglow light curves computed during three different times: before the time of the peak in the case of a uniform jet or before the transition time in the case of a structured jet ($\alpha_{\text{rise}}$, dotted lines), the part between the peak time or transition time and the jet-break ($\alpha_{\text{intermediate}}$, dashed lines), and the post-jet-break decay ($\alpha_{\text{decay}}$, dot-dashed lines). They are plotted at different viewing angles, corresponding to the colours in the colour bar on the right. A black dashed line marks a flat slope with $\alpha$ = 0. Also shown are the FXTs that are fit by a power law with temporal index $\alpha$ (red open circles) and the FXTs that are fit by a broken power law with temporal indices $\alpha_{1}$ (red points) and $\alpha_{2}$ (purple points) before and after the break, respectively. 
}
\label{fig:slopes}
\end{figure*}

As reported in \citet{2022QuirolaVasquez,2023QuirolaVasquez}, the FXT light curves can either be fit by a power law with temporal index $\alpha$ (XRT~140327, XRT~151121, XRT~161125, and XRT~191223) or by a broken power law with temporal indices $\alpha_{1}$ and $\alpha_{2}$ before and after the temporal break, respectively (XRT~030511, XRT~041230, XRT~080819,  XRT~100831, XRT~110919, XRT~141001, XRT~140507, XRT~150322, XRT~170901, XRT~191127, and XRT~210423). XRT~000519 and XRT~110103 were not well-fit fit by either a power law or a broken power law due to the presence of flares in their light curves \citep{2013Jonker, 2015Glennie, 2022QuirolaVasquez}, and are not included in the analysis of the temporal indices. We use the convention $F\propto t^{\alpha}$ regarding the sign of the temporal index.

We evaluate the temporal indices of the off-axis afterglow light curves of the 13 GRBs analysed at three different times, and we refer to the leftmost and rightmost panels of Fig.~\ref{fig:slopes} for a graphic illustration. In the case of a uniform jet (leftmost panel of Fig.~\ref{fig:slopes}), we evaluate them from the moment the light curve appears above the flux limit until the time of the peak in the light curve ($\alpha_{\text{rise}}$), between the time of the peak and the time of the jet break ($\alpha_{\text{intermediate}}$), and from the time of the jet break until the time when the light curve falls below the flux limit ($\alpha_{\text{decay}}$). In general, the jet break is visible as a steepening in the afterglow light curve when the beaming angle $\sim\Gamma^{-1}$ of the decelerating jet (where $\Gamma$ is the Lorentz factor) becomes larger than the core angle $\theta_{\rm C}$ of the jet (e.g. \citealt{1999SariPiranHalpern, Zhang}).
In the case of a structured jet (rightmost panel of Fig.~\ref{fig:slopes}), an additional time-scale is involved; it is the transition time $t_{W}$ when an observer at viewing angle $\theta_{\text{obs}} > \theta_{\rm W}$ enters the beaming cone of the near edge of the jet (see eq. (37) of \citet{2020Ryan} for the expression of $t_{W}$). We thus calculate the temporal indices before and after the transition time $t_{W}$ rather than the peak time $t_{\text{peak}}$, such that $\alpha_{\text{rise}}$ is defined as the slope of the light curve before the transition time  in the case of a structured jet. $\alpha_{\text{intermediate}}$ is then defined as the slope between $t_{W}$ and $t_{\text{break}}$, while the post-jet break temporal index $\alpha_{\text{decay}}$ remains the same as in the uniform-jet case. It should be noted, however, that the time of the jet break $t_{\text{break}}$ depends on the jet structure and viewing angle; we refer to eq. (4) of \citet{2010VanEerten} for the expression used to calculate $t_{\text{break}}$ in the case of a uniform jet, and to eq. (39) of \citet{2020Ryan} for $t_{\text{break}}$ in the case of a structured jet.

Figure ~\ref{fig:slopes} compares the temporal indices of the FXT light curves (red and purple points) to those of the off-axis version of our best-fitted 13 GRB afterglows at different viewing angles (colour-coded dashed, dotted, dot-dashed lines) up to a ratio of $\theta_{\text{obs}}/\theta_{\rm C}$ = 3\footnote{We recall that no GRB afterglows would be detected above the assumed flux limit at larger viewing angles.}. We first consider the rising temporal indices in the case of a uniform jet (second panel of Fig.~\ref{fig:slopes}) and in the case of a structured jet (third panel of Fig.~\ref{fig:slopes}). In both cases, the temporal indices $\alpha_{1}$ of a few FXTs that are fit by a broken power law can match the temporal indices $\alpha_{\text{rise}}$ of the afterglows (grouped by the curly bracket and indicated with (1) in the plot); these are XRT~141001, XRT~140507, XRT~191127, and, if we consider the structured-jet models, also XRT~080819. However, except for XRT~191127 (which has the steepest rise in its light curve compared to the other FXTs in our sample), the indices $\alpha_{1}$ and $\alpha_{\text{rise}}$ are only consistent if we are sufficiently far off-axis ($\theta_{\text{obs}}/\theta_{\rm C}\gtrsim$~1.8 in the uniform-jet case and $\theta_{\text{obs}}/\theta_{\rm C}\gtrsim$ 2.8 in the structured-jet case). 

We then consider the intermediate temporal indices (grouped by the curly bracket and indicated with (2) in the plot). In the case of a structured jet, a flat segment appears in the light curve if the line of sight lies either within the wings of the jet ($\theta_{\rm C}~<~\theta_{\text{obs}}~<~\theta_{\rm W}$) or slightly outside of them ($\theta_{\text{obs}} \gtrsim \theta_{\rm W}$). In the former case, an off-axis observer first receives emission from the jet's fainter structured wings, corresponding to an initial decline in the light curve. As the jet decelerates, the beaming cone of the brighter core of the jet now also encompasses the line of sight, resulting in a plateau in the light curve. In the latter case, the off-axis observer first sees a rise (rather than a decline) prior to the plateau, as the beaming cone of emission originating from the edge of the jet at $\theta_{\rm W}$ starts to include the line of sight at time $t_{W}$ (see the middle and right panel of fig. 3 of \citet{2020Ryan} for a schematic illustration of the two cases). Once the jet has decelerated enough to fully capture its extent, marking the time of the jet break, the light curve decays in the same way as the afterglow of an on-axis GRB.

For the structured-jet models, we may thus expect a flat segment if $\theta_{\text{obs}}\gtrsim\theta_{\rm W}$ = 2$\theta_{\rm C}$. Indeed, we find that for viewing angles in the range $\theta_{\text{obs}}/\theta_{\rm C} = 2.2-3$  the off-axis afterglow light curves show such a shallow temporal index in a range $\alpha_{\text{intermediate}}\approx -0.4-0$ that are consistent with the slopes $\alpha$ or $\alpha_{1}$ of most FXTs (see the third panel of Fig.~\ref{fig:slopes}). Exceptions are XRT 030511, XRT 041230, XRT 141001, XRT 140507, and XRT 191127, which have rising segments in their light curves.
For the uniform-jet models, there is such a match between $\alpha$ and $\alpha_{\text{intermediate}}$ only in three cases, namely XRT 151121, XRT 161125, and XRT 191223, and only when $\theta_{\text{obs}}/\theta_{\rm C}\approx$ 2. In this case, the shallow indices $\alpha_{\text{intermediate}}$ at highly off-axis angles are attributed to the fact that we are left with only the tip of the light curve that is still detectable above the flux limit (which causes the flattening in all temporal indices $\alpha_{\text{rise}}$, $\alpha_{\text{intermediate}}$, and $\alpha_{\text{decay}}$ with increasing viewing angle when far enough off-axis; see Fig.~\ref{fig:slopes}).

The comparison to the post-break temporal indices $\alpha_{2}$ of the FXTs that are fit by a broken power law shows that most FXTs have a steep decay that is largely comparable to the post-break indices $\alpha_{\text{decay}}$ of the afterglow of either a uniform or structured jet. The exceptions are XRT 191127 and XRT 210423, whose post-break decays are much steeper ($\alpha_{2} = -2.5 \pm 0.3$ and $\alpha_{2} = -3.8 \pm 1.2$, respectively), although the uncertainty on the latter value is quite large. 
We note that the post-break indices are not very sensitive to the viewing angle or the structure of the jet, as shown in the central plots of Fig.~\ref{fig:slopes}, although there is a slightly wider range of post-break indices allowed by the structured jet and they are shallower than those of the uniform jet. This can be explained as follows: since the afterglows are starting to disappear below the flux limit at $\theta_{\text{obs}}/\theta_{\rm C}\approx$ 2--3, the light curve is only detectable for a brief time after the jet break, such that it has not yet steepened to the final (asymptotic) value of the true post-jet break decay index. Since the jet break is smoother in the case of a structured jet compared to a uniform jet (e.g. \citealt{2021Lamb}), the decay right after the jet break is less steep in the case of a structured jet. We have checked that this is indeed the case in the afterglows of our GRB sample at $\theta_{\text{obs}}/\theta_{\rm C}\approx$ 2--3, and additionally, we found the post-jet break decay indices of the uniform- and structured-jet models to reach the same final value at much later times after the afterglow has disappeared below the flux limit.

Nevertheless, the light curve behaviour of FXTs that are fit by a broken power law can only be consistent with that of the GRB afterglows if both the temporal indices $\alpha_{1}$ and $\alpha_{2}$ match the temporal indices $\alpha_{\text{intermediate}}$ and $\alpha_{\text{decay}}$ at the same viewing angle. This is not the case for any FXT, as the post-jet-break decay becomes too shallow to match the (often steep) post-break decay in the FXT light curves as we go farther off-axis. 

\subsection{Optical afterglows}

\begin{figure*}
	\includegraphics[width=\columnwidth]{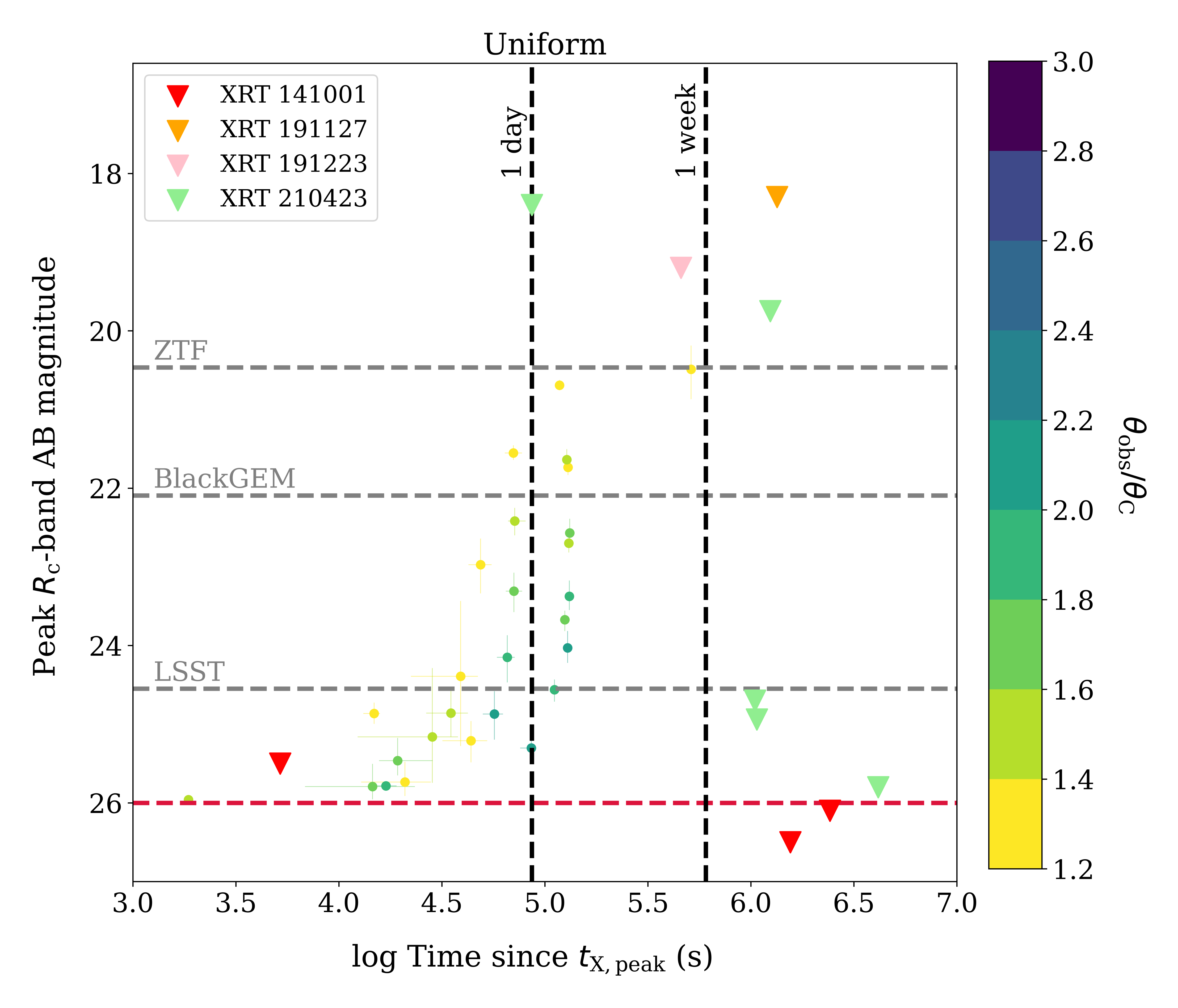}
	\includegraphics[width=\columnwidth]{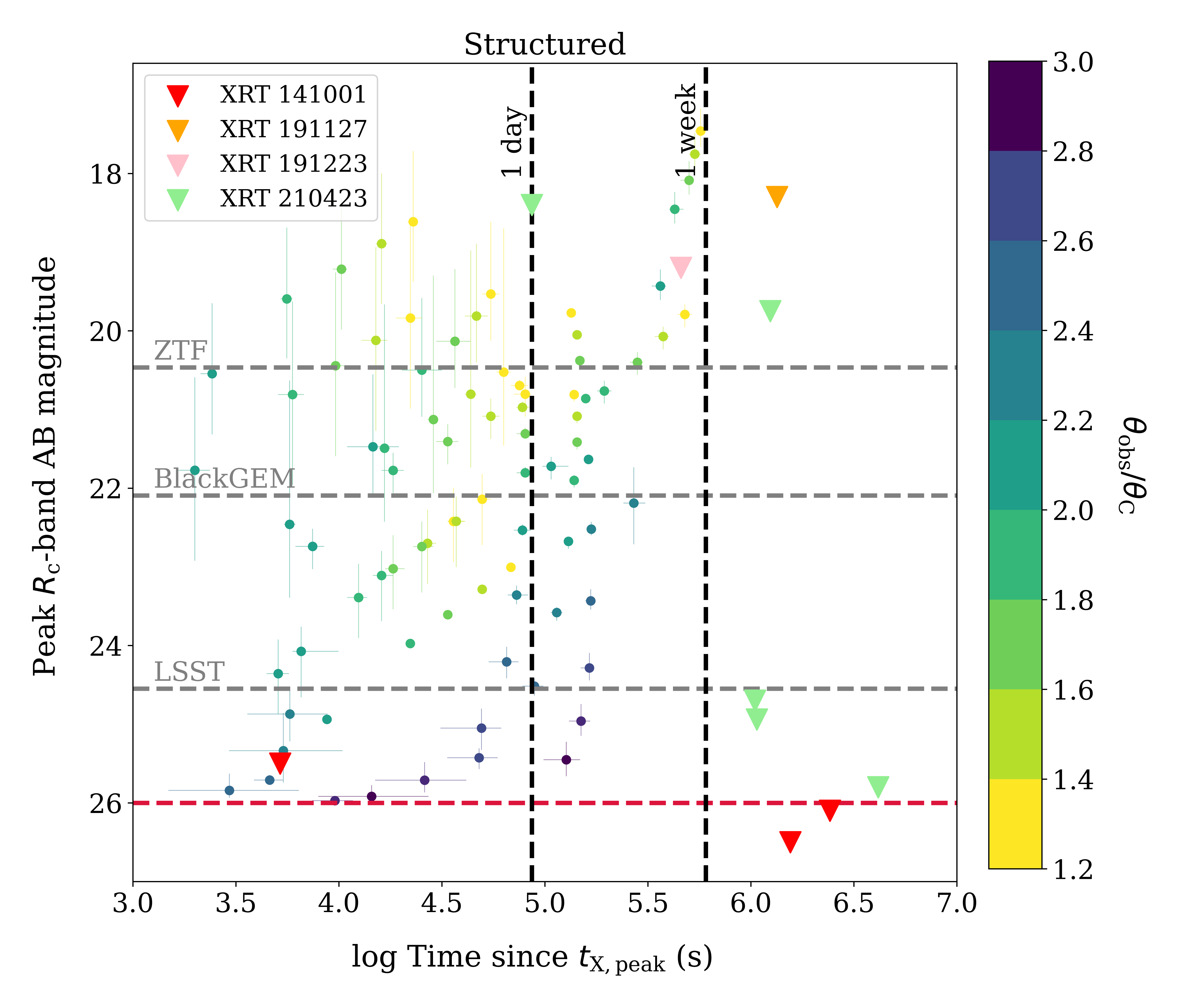}
\caption{Peak $R$-band AB magnitude of the optical afterglows of off-axis merger-induced GRBs in the sample seen from different viewing angles $\theta_{\text{obs}}/\theta_{\rm C}$ (coloured points), as a function of the time between the peak of the X-ray afterglow and the time when the optical afterglow vanishes below the magnitude limit $m_{\text{lim}}$ = 26 (red dashed line). Coloured triangles denote \textit{VLT}/VIMOS, \textit{VLT}/FORS2, and \textit{Gemini}/GMOS-S $R$-band upper limits of XRT 141001 \citep{2014TelLuo,2014TelTreistera,2014TelTreisterb,2017Bauer} and \textit{ZTF} $R$-band upper limits of XRT 191127, XRT 191223, and XRT 210423 \citep{2023QuirolaVasquez}, and additional upper limits for XRT 210423 measured by \textit{VLT}/FORS2, \textit{GTC}/HiPERCAM, \textit{Hale Telescope}/WaSP, and \textit{LBT}/LBC \citep{2021bAndreoni, 2021Rossi, 2023Eappachen, 2024QuirolaVasquez}. Grey dashed lines mark the $R$-band limiting magnitudes of \textit{ZTF} ($m_{\text{lim}}$ = 20.5), BlackGEM ($m_{\text{lim}}$ = 22.1) and \textit{LSST} ($m_{\text{lim}}$ = 24.5). Results on the left correspond to the uniform-jet models of the GRB sample, while those on the right correspond to the structured-jet model with $\theta_{\rm W}$ = 2$\theta_{\rm C}$. We note that the four GRBs with large uncertainties in their peak magnitude in the case of a structured jet with $\theta_{\text{obs}}\leq\theta_{\rm W}$ correspond to those four GRBs that have large uncertainties in their peak X-ray luminosity as well (see Sect.~\ref{sec:X-rays}).}
\label{fig:optical}
\end{figure*}

We now consider the off-axis merger-induced GRB afterglows in the optical regime. It is interesting to investigate the optical predictions for two reasons: firstly, while most FXTs in our sample lack optical/NIR upper limits close to the time of the \textit{Chandra} detection in X-rays due to the nature of their discovery in archival data, there are a few upper limits available for comparison up to a few days after the X-ray detection \citep{2017Bauer,2022QuirolaVasquez,2023QuirolaVasquez}. \citet{2023QuirolaVasquez} examined photometric archival data from the Zwicky Transient Facility (ZTF; \citealt{2019Bellm,2019Graham,2019Masci}), the Asteroid Terrestrial-impact Last Alert System (ATLAS; \citealt{2018Tonry}), the \textit{unTimely Catalog} of the  Wide-field Infrared Survey Explorer (WISE; \citealt{2010Wright,2023Meisner}), and the DESI Legacy Imaging Surveys DR10 \citep{2019Dey}, and found three FXTs for which observations both before and after the X-ray detection were obtained by \textit{ZTF} and \textit{ATLAS}. For one of them (XRT 210423), deep optical imaging with FORS2 on the 8m Very Large Telescope (VLT), HiPERCAM on the 10.4m Gran Telescopio Canarias (GTC) \citep{2023Eappachen}, the Wafer-Scale Imager for Prime (WaSP) on the 200-inch Hale Telescope at Palomar Observatory \citep{2021bAndreoni}, and the Large Binocular Cameras (LBC) on the 8.4m Large Binocular Telescope (\citep{2021Rossi} was obtained in addition. We aim to determine whether these non-detections are consistent with the predicted brightness of the optical afterglows of the off-axis GRBs in our sample, or whether they are constraining enough to possibly rule out an off-axis afterglow scenario. 

Secondly, if FXTs could indeed be the X-ray afterglows of off-axis merger-induced GRBs, predictions of the behaviour of the optical afterglows -- their optical counterparts -- are useful for potential follow-up observations of FXTs to be discovered in the future. Our goal is to predict for how long the optical afterglows of off-axis GRBs are detectable after the observed X-ray afterglow reaches its peak, if the X-ray afterglow is detectable by \textit{Chandra}.

Similar to what we did for the off-axis afterglow light curves in X-rays, we create a sample of off-axis afterglows in the optical band for the 13 GRBs analysed, by transforming the flux density light curves at an observed optical frequency of 4.634$\times 10^{14}$ Hz to the light curves in the $R_{C}$-band for different viewing angles, ranging from $\theta_{\text{obs}}/\theta_{\rm C} = 1.2$ to $\theta_{\text{obs}}/\theta_{\rm C} = 3$. We then calculate the peak AB magnitude and duration, the latter is defined as the time between the moment the X-ray flux reaches its peak and the moment the optical afterglow drops below a limiting magnitude of $m_{\text{lim}}$ = 26. 

Figure~\ref{fig:optical} shows the peak AB magnitude as a function of the logarithm of the time passed after the peak in X-rays. The results for the uniform-jet model are shown in the left panel of this figure, while the results for the structured-jet model with $\theta_{\rm W}$ = 2$\theta_{\rm C}$ are shown in the right panel.  The grey dashed lines denote the limiting magnitudes\footnote{Conversion of the $r$-band limiting magnitudes to the $R$-band ones follows the transformations by Lupton (2005): \url{http://www.sdss3.org/dr8/algorithms/sdssUBVRITransform.php\#Lupton2005}.} of \textit{ZTF} \citep{2019Bellm}, BlackGEM \citep{2016Bloemen}, and \textit{LSST} (the Rubin Observatory Legacy Survey of Space and Time; \citealt{2019Ivezic, 2022Bianco}). In the case of a uniform jet, most of the afterglows in the sample disappear below $m_{\text{lim}}$ after one day from the X-ray peak, and are detectable out to viewing angles of $\theta_{\text{obs}}/\theta_{\rm C}\lesssim$ 2 before that. They are not detectable by \textit{ZTF} at any viewing angle, but could be detectable by BlackGEM and \textit{LSST} up to one week after the light curve peaks in X-rays if they are observed only slightly off-axis. Like in X-rays, the optical afterglows of off-axis GRBs are brighter in the case of a structured jet compared to the case of a uniform jet at the same viewing angle (at least within $\theta_{\rm W}$), such that they might be detectable by \textit{ZTF}, BlackGEM and \textit{LSST} out to viewing angles $\theta_{\text{obs}}/\theta_{\rm C}\lesssim$ 2, 2.2, and 2.8, respectively. They are detectable above $m_{\text{lim}}$ out to viewing angles $\theta_{\text{obs}}/\theta_{\rm C}\lesssim$ 2.8, up to a week after the peak in X-rays for events with $z \leq$ 2.211. The predicted optical off-axis afterglows are in most cases too faint to be detectable for the survey telescope array GOTO (The Gravitational-wave Optical Transient Observer; \citealt{2022Steeghs}) given its limiting $R$-band magnitude $m_{\text{lim}}$  = 18.4 \citep{2022Steeghs}.

XRT 191127, XRT 191223, and XRT 210423 have ZTF $r$-~band upper limits of 18.3, 19.2, and 18.4 mag at 15.5, 5.3, and 1.0 days after the trigger, respectively; these are shown as coloured triangles in Fig.~\ref{fig:optical}. XRT 141001 (CDF-S XT1) has three exceptionally deep $R$-band upper limits (shown as red triangles) and one \textit{HST} F110W-band upper limit at 0.06 days, 18 days, 27 days, and 111 days after the detection in X-rays of $m_{R}$ = 25.5, 26.1, 26.5 and $m_{F110W}$ = 28.4 mag, respectively \citep{2014TelLuo,2014TelTreistera,2014TelTreisterb,2017Bauer}. From Figure ~\ref{fig:optical} it is clear that in both the uniform and structured-jet cases, the non-detections of optical counterparts of XRT 191127, XRT 191223, and XRT 210423 are consistent with the faint magnitudes and short lifetimes expected for optical off-axis GRB afterglows. An exception is XRT 141001, whose deep non-detections seem to be at odds with the expected brightness of the optical off-axis afterglows of the GRBs in our sample in the case of a structured jet with $\theta_{\rm W}/\theta_{\rm C}=2$. They would imply a jet observed quite far off-axis, with $\theta_{\text{obs}} > 3\theta_{\rm C}$.

\begin{table*}
\centering
\caption{Consistency between light curve properties of FXTs and of off-axis GRB afterglows.}
    \begin{tabular}{cccccc} \hline
     FXT & & Values of $\theta_{\text{obs}}/\theta_{\rm C}$ required to have consistent:   &  &  & Combined consistency \\ 
     & $L_{X,\text{peak}}$ & $t$ & $\alpha_{\text{(1)}}$ & $\alpha_{\text{2}}$ &  \\ \hline
    XRT 000519 &  1.6, 1.8, 2.2 & 1.4--2.8 & - & - & \xmark\\
    XRT 030511 &  1.2, 1.8, 2.2 & 1.6--2.4, 2.8 & \xmark & 1.6--2.2 & \xmark \\
    XRT 041230 &  2.8 & 1.2--2.6 & \xmark & \xmark & \xmark\\
    XRT 080819 &  2.0--2.4 & 1.4--2.8 & 3.0 & 1.2--2.2 & \xmark\\
    XRT 100831 & 1.6--2.2 & 1.6--2.4, 2.8 & 2.8--3.0 & \xmark & \xmark\\
    XRT 110103 & 2.4--2.8 & 1.2--2.6 & - & - & \xmark\\
    XRT 110919 & 2.0--2.4 & 1.2--2.8 & 2.8--3.0 & 1.2--1.6 & \xmark \\
    XRT 140327 & 2.2--2.6 & 1.2--2.8 & 2.6 & - & \checkmark ($\theta_{\text{obs}}/\theta_{\rm C}$ = 2.6) \\
    XRT 141001 & 1.2--2.0 & 1.2--2.8 & 2.8--3.0 & 1.6--2.2 & \xmark \\
    XRT 140507 & 2.2--2.6 & 1.8--2.0, 2.4, 2.8& 2.8--3.0 & 1.2--1.6 & \xmark \\
    XRT 150322 & 2.0--2.4 & 1.2--2.8 & 2.6--3.0 & 1.2--1.6 & \xmark\\
    XRT 151121 & 1.4--2.2 & 2.0--2.4 & 2.2--2.6 & - & \checkmark ($\theta_{\text{obs}}/\theta_{\rm C}$ = 2.2) \\
    XRT 161125 & 1.2--2.0 & 2.0--2.4 & 2.2, 2.4 & - & \xmark\\
    XRT 170901 &  1.2--1.4 & 2.0--2.4 & 2.6--3.0 & 1.2--2.0 & \xmark\\
    XRT 191127 &  2.0 & - & 2.4--2.8 & \xmark & \xmark \\
    XRT 191223 &  1.4--2.2 & 2.0--2.4 & 2.2, 2.4 & - & \checkmark ($\theta_{\text{obs}}/\theta_{\rm C}$ = 2.2)\\
    XRT 210423 &  1.2, 1.4, 1.8, 2.2 & 1.2--2.8 & 2.6--2.8 & \xmark & \xmark \\
    
    \end{tabular}
\tablefoot{The table shows the range of viewing angles $\theta_{\text{obs}}/\theta_{\rm C}$ for which the peak X-ray luminosity (second column), duration (third column), temporal index $\alpha$ or first temporal index $\alpha_{1}$ if the FXT is fit by a broken power law (fourth column), and second temporal index $\alpha_{2}$ (fifth column) of each FXT are consistent with individual off-axis GRB afterglows (in case of a structured jet only) at the same viewing angle. Regarding the temporal indices, a crossmark indicates that there is no match between the relevant temporal index of the FXT with those of the GRB afterglows, while a dash indicates that the FXT was fit by a simple power law and thus only has one temporal index $\alpha$. XRT 000519 and XRT 110103 were not considered in the analysis of the temporal indices because they could be fit by neither a simple power law nor a broken power law. The sixth column denotes whether the FXT is consistent with the GRB afterglows \textit{in all three properties} (luminosity, duration, and temporal indices), and if so, at which viewing angles.}
\label{tab:FXT-compatible}
\end{table*}

\section{Discussion}
\label{sec:Discussion}

When comparing the light curve properties of both samples as done in Sec.\ref{sec:Results}, it is crucial to look for viewing angles for which all three properties analysed, namely the luminosity, duration, and temporal indices, are consistent at the same viewing angle (which may differ between the sources) in order for an off-axis afterglow scenario to fit the FXT data in a fully self-consistent manner. For example, although the high peak X-ray luminosity of XRT 141001 ($\approx 10^{47}$ erg s$^{-1}$) calls for a more on-axis viewing angle, the optical non-detections imply that it was observed highly off-axis ($\theta_{\text{obs}} > 3\theta_{\rm C}$), and together with a short duration ($\approx$ 5 ks), a shallow pre-break rise ($\alpha_{1}$ = 0.4) and post-break decay ($\alpha_{2}=-1.6$), its properties are not self-consistent for any viewing angle. 
The constraints coming from the comparison of the temporal indices tell us that the best viewing angle to match the off-axis afterglows of our selected GRBs is between 2.2$\theta_{\rm C}$ and 3$\theta_{\rm C}$ in the structured-jet case, essentially to reproduce the plateau segments observed in those FXTs best-fit by a broken power law. 
The durations of the off-axis GRB afterglows in our sample are too short to be compatible with those of FXTs in the uniform-jet case, while there is a larger overlap in the structured-jet case, such that we can reproduce the FXT durations with observing angles in the range (2--3)$\theta_{\rm C}$.
When comparing the peak X-ray luminosity of the two populations, we found that the distributions of FXTs and the afterglows are consistent for only one inclination ($\theta_{\text{obs}} = 2.2\theta_{\rm C}$), provided the jet is structured and has a truncation angle $\theta_{\rm W} = 2\theta_{\rm C}$.
This implies that the chances of having all three properties being consistent with one another are maximised for only one viewing angle, $\theta_{\text{obs}} = 2.2\theta_{\rm C}$, in the case of a structured jet. However, while this holds when comparing the distributions of the properties of FXTs with those of GRB afterglows, this does not exclude potential further matches at different viewing angles if one considers individual events, rather than the distribution of the whole population. In fact, the different intrinsic luminosities of the GRB afterglows analysed make the corresponding off-axis afterglows span a quite wide range of peak X-ray luminosities for the same viewing angle (see for example the yellow points corresponding to 1.2$\theta_{\rm C}$ which span the range of luminosities $\sim 10^{46-49}$erg/s).  
Indeed, evaluating individual FXTs and comparing their observational properties with the ones of the sample of off-axis GRB afterglows can reveal favourable matches also beyond the unique value of 2.2$\theta_{\rm C}$, as can be seen in Table~\ref{tab:FXT-compatible} (in the following, we will focus on the case of a structured jet with truncation angle $\theta_{\rm W} = 2\theta_{\rm C}$, since the uniform-jet case holds the least chances of unifying all the properties of FXTs and GRB afterglows). For example, FXTs whose X-ray light curve is well-fit by a single power law, namely XRT~140327, XRT~151121, XRT~161125, and XRT~191223, have temporal indices ($\alpha = -0.2, -0.3,-0.5,-0.4$, respectively) consistent with those of off-axis afterglows specifically for viewing angles $\theta_{\text{obs}} = (2.2-2.4)\theta_{\rm C}$ (XRT~161125 and XRT~191223), $\theta_{\text{obs}} = (2.2-2.6)\theta_{\rm C}$ (XRT~151121), and $\theta_{\text{obs}} = 2.6\theta_{\rm C}$ (XRT~140327), as can also be seen in Fig.~\ref{fig:slopes}. For XRT~140327, its luminosity ($L_{X,\text{peak}} = 6.3\times 10^{44}$~erg s$^{-1}$) and duration (13.4 ks) are consistent with the luminosity and duration distribution of off-axis afterglows at the same viewing angle. At $\theta_{\text{obs}}$ = 2.2$\theta_{\rm C}$, the peak X-ray luminosities and durations of XRT~151121 as well as XRT~191223 are also consistent with the luminosity and duration of GRB afterglows. On the contrary, while the temporal index of XRT~161125 calls for viewing angles $\theta_{\text{obs}} = (2.2-2.4)\theta_{\rm C}$, at these viewing angles the luminosity distributions of the afterglows cannot accommodate its high luminosity at its assumed redshift ($L_{X,\text{peak}} = (1.9\pm 0.8)\times 10^{47}$~erg s$^{-1}$). 
To summarise, although it may be difficult to accommodate the FXTs with relatively high peak X-ray luminosities and short durations in our sample, there is a limited range of viewing angles for which some FXTs have temporal slopes, luminosities and durations compatible with those of off-axis GRB afterglows analysed in this work.

However, there are a few caveats to take into account when interpreting the results described above. The first one regards selection effects in our sample of GRBs. In fact, our analysis is based on only 13 GRBs, whose afterglow light curves have been required to be well-sampled by broad-band observations, in order to constrain the afterglow parameters as much as possible. This high-cadence multi-wavelength requirement could result in a selection of particularly bright GRBs at the expense of the fainter ones, which are less likely to have detections or even be followed up while their emission fades away. The peak energy fluxes measured in 1-s bins in the 15-150 keV energy range of our selected sample of GRBs are in the range $0.33-5.2 \times 10^{-7} \rm erg \, cm^{-2} \, s^{-1}$. Compared to the peak energy flux values in the third Swift catalogue (\citealt{Lien2016}, see e.g. their fig. 18), these values suggest that our selected sample of GRBs is a bright subset of the whole population of Swift-detected merger-induced GRBs. In addition, their prompt $\gamma$-ray fluences in the same band lie in a range $3.0\times10^{-8}$--$1.4\times10^{-6}$ erg cm$^{-2}$ with a median value of  3.4$\times10^{-7}$ erg cm$^{-2}$, making them moderately bright compared to other short GRBs in the sample by \citealt{Lien2016} (e.g. their fig. 14). Moreover, most of the 13 GRBs in our sample have been detected by Swift, whose detections are dominated by the intrinsically more luminous events, rather than the fainter ones, and of course there is a population of low-luminosity GRBs towards which we are less sensitive or even blind. 

This also prompts the question about the possible existence of a larger region in the luminosity-duration plane where these lower-luminosity GRB afterglows could be consistent with the FXTs, if one allows the intrinsic luminosity to vary. Given the region in Fig. \ref{fig:X-rays} currently filled by the sample of 13 GRBs analysed in this work, one could wonder if filling this plot with events a factor 10 or 100 less luminous would make them more similar to the current sample of FXTs, both in terms of luminosity and duration. 
However, Figure~\ref{fig:X-rays} shows that our current sample of afterglows already starts to vanish below the flux limit at $\theta_{\rm obs} \geq 2.2\theta_{\rm C}$, such that it is unlikely that a population with a luminosity 10~--~100 times lower will be detectable at greater off-axis angles. The lack of detectable afterglows at greater off-axis angles has a major effect on the consistency of the temporal indices. Table~\ref{tab:FXT-compatible} shows that those FXTs that cannot match the population of off-axis afterglows in terms of all three properties (luminosity, duration, and temporal indices) all require $\theta_{\rm obs} \geq 2.2\theta_{\rm C}$ to match the temporal index $\alpha_{(1)}$ of the plateaus in their light curves.

We have simulated a lower luminosity population of GRB afterglows by decreasing the inferred isotropic-equivalent energy $E_{0}$ of each GRB in our sample by a factor 10 (although we note that the peak X-ray luminosity does not only depend on the jet parameter $E_{0}$; see for example the expression in \citealt{2020Oconnor}, the full exploration of the parameter space of the afterglow model is, however, outside the scope of this work). The results for the structured-jet case with $\theta_{\rm C} = 2\theta_{\rm W}$ are shown in the left panel of Fig.~\ref{fig:X-rays-III} in Appendix~\ref{sec:appendix-alt}. Indeed, the lower luminosities of a sample of off-axis GRB afterglows can match the luminosities of some individual FXTs well. However, these less energetic afterglows do not give rise to a plateau in the light curve for relevant off-axis viewing angles, making the light curves inconsistent with those of the observed FXTs. 
In fact, there are only two GRBs that remain detectable at $\theta_{\rm obs} \geq 2.2\theta_{\rm C}$ and none remaining beyond $\theta_{\rm obs} = 2.6\theta_{\rm C}$. 
Moreover, the durations of the off-axis sub-energetic GRBs are shorter than the durations of FXTs, with the best match of the two distributions happening at the smallest viewing angles ($1.2-1.6 \, \theta_{\rm obs}/\theta_{\rm C}$).

We also tested the hypothesis that a better chance to match the temporal indices of our population of FXTs comes from considering more off-axis angles (between $2.2\theta_{\rm C}$ and $4\theta_{\rm C}$). For this purpose, a more luminous population of GRB afterglows would be needed: we consider this scenario now by increasing the isotropic-equivalent energy $E_{0}$ by a factor of 10. The results (again for the structured-jet case with $\theta_{\rm C} = 2\theta_{\rm W}$) are shown in the right panel of Fig.~\ref{fig:X-rays-III}. In this case there is an increased overlap in the distribution of the peak X-ray luminosity of the two populations at greater off-axis angles, which would also make the temporal indices consistent. However, a more luminous population of GRB afterglows is also significantly longer lived above a given flux limit: this effect is visible from the plot, where the durations of the simulated brighter GRBs afterglows are much longer than what previously found,  drastically reducing the overlap with the distribution of the FXT durations.
Furthermore, the current range of isotropic-equivalent kinetic energies for our sample $E_{0}\approx 10^{50-53}$~erg (see Appendix~\ref{sec:appendix-fitting}) is already quite high, and an increase of $E_{0}$ by a factor ten or more would push the kinetic energy of the GRBs to extreme values.

There is also an additional uncertainty on both populations' inferred peak X-ray luminosities that needs to be taken into account, and it is due to the lack of robust redshift estimates (nine out of 17 FXTs and four out of 13 GRBs have an uncertain redshift). This emphasises the need for deep searches for potential host galaxy candidates in order to obtain photometric redshifts and to probe the true energetics of FXTs and merger-induced GRBs.

The above considerations call for a deeper investigation of the possible connection between GRBs afterglows and FXTs. For example, this could be achieved via a more detailed analysis offered by population studies of merger-induced GRBs, taking into account the whole range of intrinsic luminosities and possibly also of afterglow properties, which offers a promising route towards a more complete exploration of the parameter space. Such population studies for GRBs have been developed in the community (see e.g. \citealt{DAvanzo2014,Wanderman2015,Ghirlanda2015} and the recent study by \citet{Salafia2023} taking into account selection effects). We compare some of the main results of our study with those of the population study on orphan GRBs presented in \citealt{Ghirlanda2015}. Their figure 3 shows the duration of the off-axis afterglow as a function of the X-ray limiting flux sensitivity: if we assume the same Chandra flux limit used in this work ($3\times10^{-14} \rm erg/s/cm^2$, corresponding to $\sim 10^{-6}$ mJy), the durations are expected to be $\sim \rm log(t [s])=5.08^{+0.61}_{-0.54}$  (using their fitting formula in their table 1). These are consistent with the GRB off-axis durations we found in our analysis (median of $\rm log(t [s]) = 4.72^{+0.18}_{-0.44}$ for the uniform jet, and $\rm log(t [s]) = 4.47^{+0.38}_{-0.51}$ for the structured jet), and they are inconsistent with the much shorter FXT durations (median of $\rm log(t [s]) = 3.81^{+0.36}_{-0.23}$, see Figure \ref{fig:X-rays}).
These population studies could also be exploited to gain insights from the comparison of the detection rates of the orphan afterglows and those of FXTs. As pointed out in \citet{2022QuirolaVasquez}, the volumetric density rate of FXTs is quite high and barely consistent with the peak of the rate of short GRBs (also when accounting for different beaming corrections). 
The FXT rates reported in \citet{2022QuirolaVasquez} are more consistent with the rate of long GRBs, in particular with the low-luminosity fraction of them, and this promising channel could be investigated through an analysis similar to what is presented in this work. A caveat of this scenario is that long GRBs typically have smaller host offsets than short GRBs (e.g. \citealt{2022Fong}). For example, the physical offsets of FXTs 8, 9, 16, 19, 21, and 22 are consistent with those of short GRB hosts, while less than 15\% of long GRB hosts have similar or larger offsets \citep{2023QuirolaVasquez}.
Another alternative approach to the one applied in this work would be to directly fit the \textit{Chandra} FXT light curves with an afterglow model (e.g. as attempted in \citealt{2021Sarin}). However, the often low number of counts in \textit{Chandra} FXT observations might hamper a meaningful derivation of the constraints on the several afterglow free parameters. Furthermore, only a small subset in the whole FXT population might be of merger-origin, and already in our sample of 13 GRBs, many afterglows were found to be contaminated by non-afterglow emission components. Together, these two facts suggest that a direct fit to the currently known FXT light curves may be difficult.

When considering the possible connection between FXTs and off-axis GRB afterglows, it is crucial to take into account also the observational constraint coming from the lack of accompanying $\gamma$-ray emission. Provided there is a jet and it is responsible for an emission in $\gamma$-rays, the absence of $\gamma$-ray triggers can put stringent limits on the viewing angle, on the jet's bulk Lorentz factor, and on the luminosity (through the compactness argument, see e.g. \citealt{Lithwick2001} or the recent off-axis version in \citealt{Matsumoto2019}).
In general, the slightly off-axis perspective suggested by our results ($\theta_{\rm obs} \sim (2-3)\theta_{\rm C}$) can be compatible with the lack of $\gamma$-ray detection (as also suggested by \citealt{2018BeniaminiNakar}), and in fact, it would be the favoured scenario allowing for the prompt emission to be missed while still having the off-axis afterglow detectable in \textit{Chandra}. 
Indeed, there are many more GRBs whose jet is not pointed towards us: for typical opening angles of GRBs of a few degrees, $\theta_{\rm C} \sim 0.1$ rad, they should outnumber the population of on-axis GRBs (by a factor $\propto (1 - \rm cos \,\theta_{\rm C})^{-1} \sim 260$). These events, whose prompt gamma-ray emission is undetected, can be observed during the afterglow phase, when the jet, interacting with the external medium, starts to decelerate. In particular, when the jet has decelerated enough so that the relativistic beaming $\propto 1/\Gamma$ includes the observer's line of sight (that is, when $1/\Gamma \sim \rm sin(\theta_{\rm obs} - \theta_{\rm C})$), the off-axis observer starts to see the afterglow radiation. However, for highly relativistic jets as the ones typically observed in GRBs ($\Gamma \sim 300-500$), the emission is strongly boosted forward, and the flux is dramatically suppressed for off-axis observers (due to the relativistic Doppler factor, $\delta = (\Gamma \cdot (1-\beta \, \rm cos\theta_{\rm obs})^{-1}$). The material dominating the emission is that which just entered the beaming cone, but the related $\Gamma$ will be much smaller for more off-axis observers so that the related emission will be much less boosted, hampering the detectability of the emission.
The combination of these effects implies that the highest chance of detection for off-axis events, are those that are slightly off-axis, which is consistent with our findings. However, we stress that the result of the narrowness of the viewing angles derived in this work when comparing with FXT properties is not simply due to geometrical and beaming effects. Our results arise also from the combined constraints derived on the durations, and perhaps more importantly, on the temporal indices of the light curves. For example, assuming a deeper flux limit (see the text below and  the results in Appendix A and Fig.~\ref{fig:X-rays-IV}), in principle it would be possible to detect more off-axis afterglows at for example 4 $\theta_{\rm obs} / \theta_{\rm C}$, but their duration and temporal indices would not be compatible with the FXT properties considered in this work.

Nonetheless, the region of the parameter space allowed by missing the prompt emission but still detecting the afterglow one strongly depends on both the prompt emission properties (luminosity, peak energy, variability timescale, etc.) and the afterglow emission properties, as well as common properties like the bulk Lorentz factor, the ratio between the viewing angle and the core angle, and the structure of the jet. Therefore a proper comprehensive and quantitative assessment of the related limits needs to be carefully performed case by case (see the recent application to the case of the relatively low-luminosity GRB 190829A in \citealt{Salafia2022}). 

We note that for the inclinations ($\theta_{\text{obs}}\approx (2.2-3)\theta_{\rm C}$) at which the GRB afterglow temporal indices $\alpha_{\text{intermediate}}$ are consistent with the plateau indices ($\alpha$, $\alpha_{1}$) in the light curves of a few FXTs, the corresponding light curves are progressively fainter for larger off-axis angles and we are left with very few afterglows detectable above the assumed flux limit, hampering the statistical comparison between the two populations. For example, beyond $\theta_{\text{obs}}\gtrsim 2.6\theta_{\rm C}$ in the structured-jet case, less than five out of 13 GRBs remain. This makes one wonder what the light curves of our fitted 13 GRB afterglows would look like beyond such viewing angles, and by extension, such flux limits.
For this reason, we have repeated the analyses presented above but assuming a 10 times deeper flux limit (the nominal flux limit of \textit{Chandra}/ACIS for an on-axis point source, which is $\approx 3\times10^{-15}$ erg cm$^{-2}$ s$^{-1}$ given an exposure time of 20 ks in the 0.3--10 keV band). The resulting luminosity-duration plane is shown in Fig.~\ref{fig:X-rays-IV} in Appendix~\ref{sec:appendix-alt}. From this analysis, we obtained the following results: \\
(1) Given the deeper flux limit, both in the uniform- and structured-jet case the off-axis afterglows are detectable out to viewing angles $\theta_{\text{obs}} = 4\theta_{\rm C}$, with four GRBs remaining at viewing angles $\theta_{\text{obs}} = 3,4\theta_{\rm C}$ in the structured-jet case; \\
(2) A deeper flux limit affects both the luminosities and the measured duration, but mostly the latter; while adding a few more GRBs at luminosities fainter than $10^{44}$ erg/s (in a region of the parameter space already probed by previous results), the durations are pushed towards much longer values, up to $t \sim 10^{6.1}$~s (on average, the afterglows above the nominal \textit{Chandra}/ACIS flux limit last approximately one order of magnitude longer than the previous results); 
\\
(3) We observe a plateau ($\alpha_{\text{intermediate}}\approx -0.2$) in the afterglow light curves also at $\theta_{\text{obs}} = (3-4)\theta_{\rm C}$, with only four GRBs remaining at these viewing angles; 
\\
(4) Like the X-ray counterparts, also the optical afterglows are now detectable out to viewing angles $\theta_{\text{obs}} = (3-4)\theta_{\rm C}$. 

Summarizing, the analysis assuming the nominal flux limit for on-axis point-source detectable by \textit{Chandra} would provide us with even more separation between the two populations: the off-axis GRBs afterglows would last too long with respect to the current FXTs and the observing angles for which a plateau would be visible in the light curve do not have any compatible counterparts in the luminosity-duration plane. 
However, we stress that the afterglows evaluated above this deeper flux limit do not allow for a fair comparison to the current sample of FXTs; the actual flux limit of the observations of the FXTs reported by \citet{2022QuirolaVasquez,2023QuirolaVasquez} was not the nominal \textit{Chandra} flux limit, as this depends on several aspects of the observations, for instance, the off-axis angle of the source with respect to the optical axis of the satellite, the spectrum of the source,  and the age of the spacecraft. 
Had the flux limit been the nominal \textit{Chandra} flux limit, we would see a major effect: we would observe fainter FXTs, and consequently, FXTs of longer durations. A population of longer/fainter FXTs might still be hidden in the \textit{Chandra} archive.
The current sample of FXTs is subject to an observational bias against the detection of long-duration FXTs.
The algorithm used to detect FXTs in \textit{Chandra} archival data by \citet{2022QuirolaVasquez, 2023QuirolaVasquez} may prevent the discovery of long-duration FXTs. Once found a sufficient photon count difference within one \textit{Chandra} observation, this detection algorithm divides that light curve into 20 ks segments and additionally requires a sufficient variation in photon counts within them. If there is no variability in them, then the FXTs would not be detected. The plateaus in the light curves of off-axis merger-induced GRB afterglows -- which are required to explain the flat segments in those of FXTs -- may last over 100 ks, without exhibiting much variability over this time interval. To detect a potential long-duration FXT, a time window exceeding 20 ks is thus required. 

Moreover, there are also some caveats related to the modelling of the afterglow emission. The first one regards the structure of the jet and the truncation angle assumed. In this work, we tested the "top-hat" (uniform-jet) case and the alternative case of a structured jet, choosing the commonly used Gaussian profile and fixing a truncation angle $\theta_{\rm W}=2\theta_{\rm C}$. We also tested our results by assuming a different truncation angle, namely $\theta_{\rm W}=1.5\theta_{\rm C}$ (see Appendix \ref{sec:appendix-alt}), obtaining similar conclusions. However, a few other alternative jet structures have been suggested in the literature (see the recent review \citealt{Salafia2022review}), including a power-law structure or the more complex composition of two nested uniform jets (one narrow, fast jet surrounded by a wider, slower one), and attempts to investigate their roles in determining how the emission appears have been explored (see e.g. \citealt{2020Beniamini, 2021Lamb, 2022Beniamini}). This represents an active research field, especially after GW170817/GRB170807A (where a structured jet observed off-axis provides the most satisfactory explanation of observations), and discerning the jet structure with the current observational data, while possible in theory, proved to be difficult so far. In this work we opted for the most commonly assumed Gaussian profile, limiting the truncation angle to $\theta_{\rm W}$ = 1.5$\theta_{\rm C}$ or $\theta_{\rm W}$ = 2$\theta_{\rm C}$. However, this by no means constitutes a complete exploration of the parameter space, and we emphasise a wider range of jet structures should be probed in order to explore the possible connection between FXTs and off-axis merger-induced GRB afterglows.

The second caveat is related to the \texttt{afterglowpy} model used in this work. In fact, this basic (yet the only publicly available) afterglow model does not include an initial coasting phase of constant $\Gamma_{0}$, as the jet is assumed to be already decelerating at the beginning of the observations \citep{2020Ryan}. The coasting phase ends at the deceleration time, which is inversely proportional to the bulk Lorentz factor $\Gamma_{0}$ (e.g. \citealt{2020BeniaminiX, 2018BeniaminiNakar}). The coasting phase generally affects the light curve at early times, when the light curve is expected to rise (with a temporal index $\alpha_{\text{coast}}$ = 2, 3 depending on the observational frequency;  e.g. \citealt{2020Beniamini, 2020Ryan, 2022Fraija}) until it reaches its peak at the deceleration time. However, we found no evidence for such a rising phase in the 13 afterglow light curves fitted in this work, and the shape of the light curves observed (see Fig.~\ref{fig:fits}) suggests that most of the afterglow observations considered here have indeed been taken after deceleration has begun. 
One exception may be GRB 070714B, for which \citealt{2017Gao} suggest that the early optical afterglow could be the afterglow onset. However, we cannot reproduce the X-ray and optical afterglows of GRB 070714B with the afterglow parameters adopted by \citealt{2017Gao}. On the other hand, \citealt{2015Gompertz} suggest that the peak in the optical light curve might also be the spectral peak frequency passing through the $R$-band.
In short, we conclude that the absence of the coasting phase in the \texttt{afterglowpy} modelling is unlikely to influence our conclusions.
Another possible limitation is that \texttt{afterglowpy} does not include the effects of synchrotron self-absorption (SSA) or synchrotron self-Compton (SSC) on the afterglow. 
SSA mainly affects the early radio afterglow, as the afterglow is reduced at frequencies below the self-absorption frequency $\nu_{a}$. However, in our sample of GRBs, data in the radio band (when available) constitute mostly of upper limits (only GRB 050724 had two detections; see Fig.~\ref{fig:fits}) at $\rm t \geq 1$ day, so that our fitting results, essentially constrained by the X-ray and optical data, should not be significantly affected by the lack of SSA in the applied model. 
The SSC process occurs when electrons that produce synchrotron photons up-scatter the same photons to higher energies through inverse-Compton (IC) scattering, resulting in an increased electron cooling rate and thus lowering the X-ray afterglow above the cooling frequency (e.g. \citealt{Zhang}). The SSC process has been shown to impact the X-ray afterglow flux effectively up to $\sim$ 1 day after the burst \citep{2021Jacovich}. It gives rise to a high-energy ($>$ 100 MeV) component to the afterglow, and it has been recently demonstrated to be present in a few, bright, long GRBs \citep{MAGIC14C,Miceli2022,Salafia2022}. A high-energy spectral component was detected for two sources in our GRBs sample, namely GRB 090510 (e.g. \citealt{2010Ackermann, 2011Panaitescu, 2016Fraija, 2021Joshi}) and GRB 160821B \citep{2021Acciari}. In the case of GRB 090510, \citet{2016Fraija} concluded that the X-ray and optical fluxes are dominated by forward shock emission. For GRB 160821B, a $\gtrsim$ 0.5 TeV $\gamma$-ray signal was detected four hours after the initial burst \citep{2021Acciari}. Even though the SSC contribution can vary by orders of magnitude, depending on the inferred jet parameters, both \citet{2021Acciari} and \citet{2021Zhang} found that the calculated SSC contribution is at least one order of magnitude too low to explain the observed flux in the case of GRB160821B. The X-ray afterglow of this burst is always dominated by the forward shock emission at $t >$ 1000 s. We thus conclude that it is unlikely that contributions from SSA and SSC have affected our broadband modelling significantly. 

During the selection of the observational data of the 13 GRBs analysed, we identified data points that are unlikely to belong to the forward shock of the afterglow component; 10 out of 13 GRBs have data points that were affected by such non-afterglow emission. The prevalence of these features in the merger-induced GRB afterglows offers other possible explanations for similar features observed in the light curves of FXTs, especially their plateaus. In the context of the afterglow interpretation, they are often interpreted as a signature of a magnetar central engine (e.g. \citealt{2012Margutti, 2020Matsumoto}). Indeed, a magnetar formed in the aftermath of a BNS merger has been suggested also as a possible central engine for several FXTs in the sample exhibiting a plateau \citep{2017Sun, 2019Sun, 2019Xue, 2020DadoDar, 2021Sarin, 2021AiZhang, 2022Lin} and it represents an alternative scenario to the one investigated in this work. For GRBs, there are several interpretations of the plateau in addition to a magnetar scenario; these include a hyperaccreting black hole as a central engine \citep{2008Kumar}, evolving microphysical jet parameters \citep{2006Ioka}, significant reverse shock emission \citep{2007Genet}, explanations pertaining to the geometrical structure of the jet (e.g. \citealt{2004Huang, 2006Toma, 2020Ryan}) -- possibly observed off-axis \citep{2020BeniaminiX}, high-latitude emission \citep{2020Oganesyan, 2020Ascenzi, 2020Beniamini, 2022Ronchini}, and finally, the expansion of a jet with a low initial bulk Lorentz factor into a medium-low density wind \citep{2020BeniaminiX, 2022DereliBegue}. The latter two scenarios do not require any long-lasting activity of a central engine.
A plateau due to high-latitude emission is not visible at an arbitrary viewing angle, however, and is discernible only when the viewing angle still lies within the core of the jet \citep{2020Ascenzi}. A low initial bulk Lorentz factor ($\Gamma_{0}$) which causes a late deceleration time could cause a plateau in the afterglows of (also on-axis) GRBs, but it requires a wind-like circum-burst medium \citep{2020BeniaminiX, 2022DereliBegue} which is not the environment expected for merger-induced GRBs. 

\section{Conclusion}
\label{sec:Conclusion}
Among the possible origins of extragalactic fast X-ray transients (FXTs) presented in \citet{2022QuirolaVasquez,2023QuirolaVasquez} (including tidal disruption events, supernova shock breakouts, and BNS mergers), we investigated if their light curve properties are consistent with the X-ray afterglows of off-axis merger-induced GRBs. We selected a sample of 13 merger-induced GRBs, with well-sampled multi-wavelength afterglow light curves, and we fitted their observed emission with the semi-analytical model \texttt{afterglowpy} \citep{2020Ryan} through a Markov Chain Monte Carlo (MCMC) approach. We then transformed the fitted afterglow light curves observed on-axis into their corresponding off-axis versions, by changing $\theta_{\text{obs}}/\theta_{\rm C}$ whilst keeping the rest of the jet parameters inferred for each GRB fixed. Considering both uniform and structured jets with a truncation angle $\theta_{\rm W} = 2\theta_{\rm C}$, we compared the light curve properties (peak X-ray luminosity, duration, and temporal indices) of FXTs with those of the off-axis version of our best-fitted 13 merger-induced GRB afterglows.

Comparing the two samples, we found that a narrow range of viewing angles ($\theta_{\text{obs}}\sim (2.2-3)\theta_{\rm C}$) and a structured jet are required to reproduce the observed plateaus ($|\alpha|\lesssim$ 0.3) in the FXT light curves, which are instead missing in the case of a uniform jet.

A similar range of viewing angles ($\theta_{\text{obs}}\sim (2-3)\theta_{\rm C}$) is required to explain the $\sim$ks durations of FXTs in the structured-jet case, while the durations in the uniform-jet case are never consistent with the (much shorter lasting) FXTs at any viewing angle.
Regarding the peak X-ray luminosities, there is only one viewing angle for which both samples are consistent with one another, at $\theta_{\text{obs}}= 2.2\theta_{\rm C}$. 

This makes it difficult to accommodate all light curve properties of FXTs with an off-axis merger-induced GRB afterglow model on a population level, as some FXTs in our sample have high peak X-ray luminosities, short durations, and flat segments in their light curves which are generally difficult to reconcile at a single, unique viewing angle. Selection biases, intrinsic variations in luminosities and durations, and a lack of robust redshift estimates might play a role in this inconsistency.
Nevertheless, considering that individual FXTs exhibit diverse properties and considering the intrinsic variation in luminosities in the GRB sample, focussing on single events revealed favourable matches with off-axis afterglows at the same slightly off-axis viewing angle ($\theta_{\text{obs}}\sim (2.2-2.6)\theta_{\rm C}$) for some FXTs. 

Moreover, we exploited our best-fitting afterglow parameters to produce predictions in the optical regime for the 13 analysed GRBs, in order to compare them with the currently available upper limits on optical counterparts to FXTs and for potential follow-up of future ones.
In both the uniform- and structured-jet models, predictions of the optical afterglows of off-axis GRBs are consistent with the $R$-band upper limits that are available for three FXTs. For XRT 141001, deeper \textit{VLT} and \textit{Gemini} upper limits are available and they imply an afterglow observed at a large off-axis viewing angle $\theta_{\text{obs}} > 3\theta_{\rm C}$. If FXTs are due to a GRB afterglow observed off-axis, then depending on the jet structure and viewing angle, large ground-based optical survey telescopes such as \textit{ZTF}, BlackGEM and \textit{LSST} should be able to detect many of the merger-induced GRB afterglows in our sample out to viewing angles of $\theta_{\text{obs}} \lesssim 3\theta_{\rm C}$, and up to a week after the peak in the X-ray light curve is observed.

Despite the analysis presented in this work being based on a small sample of 13 merger-induced GRBs with a well-sampled afterglow light curve, the results show a promising - although small - region of the parameter space where the combination of luminosities, duration and temporal indices of off-axis GRB afterglows at the same viewing angle are consistent with those of FXTs.
Future observations, especially those from the recently launched Einstein Probe mission \citep{2022Yuan}, coupled with comprehensive GRB population studies and advanced afterglow models, hold the key to unveiling the possible connection between FXTs and GRBs.

\begin{acknowledgements}
This work made use of data supplied by the UK Swift Science Data Centre at the University of Leicester. We thank D. B. Malesani for helpful comments and discussions. We thank D.A. Kann for collecting and providing the GRB optical afterglow data used in this research. We acknowledge funding from: ANID Millennium Science Initiative Program ICN12\_009 (FEB, JQ-V); ANID CATA-BASAL \#FB210003 (FEB), and ANID FONDECYT Regular \#1200495 (FEB) and \#1241005 (FEB). MER acknowledges support from the research programme Athena with project number 184.034.002, which is financed by the Dutch Research Council (NWO). P.G.J.~has received funding from the European Research Council (ERC) under the European Union’s Horizon 2020 research and innovation programme (Grant agreement No.~101095973).
\end{acknowledgements}

\section*{Data Availability}
All GRB X-ray, optical, and radio afterglow data are publicly available through either the \textit{Swift} light curve Repository 
\\
(\url{https://www.swift.ac.uk/xrt_curves/}) 
or referenced literature. The code used for the broadband light curve fitting, the analysis of FXT and GRB data, and the production of figures can be found on Zenodo via \url{https://doi.org/10.5281/zenodo.12587018}.

\bibliographystyle{aa} 
\bibliography{references} 

\begin{appendix}

\section{Effects of varying the isotropic-equivalent kinetic energy $E_{0}$, the flux limit, the truncation angle $\theta_{\rm W}$, and the integration time $\tau$}
\label{sec:appendix-alt}

\begin{figure*}[!ht]
	\includegraphics[width=\columnwidth]{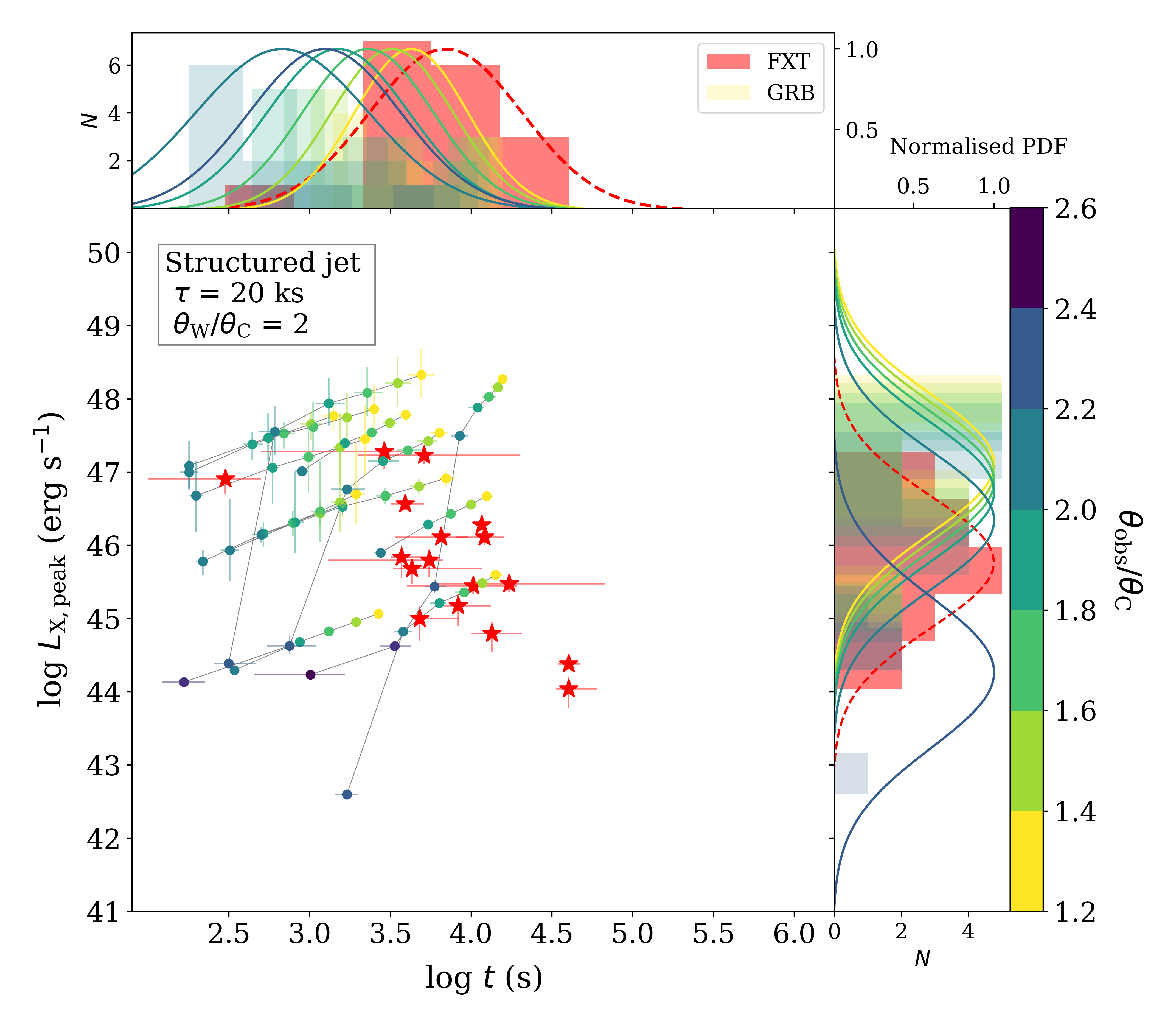}
	\includegraphics[width=\columnwidth]{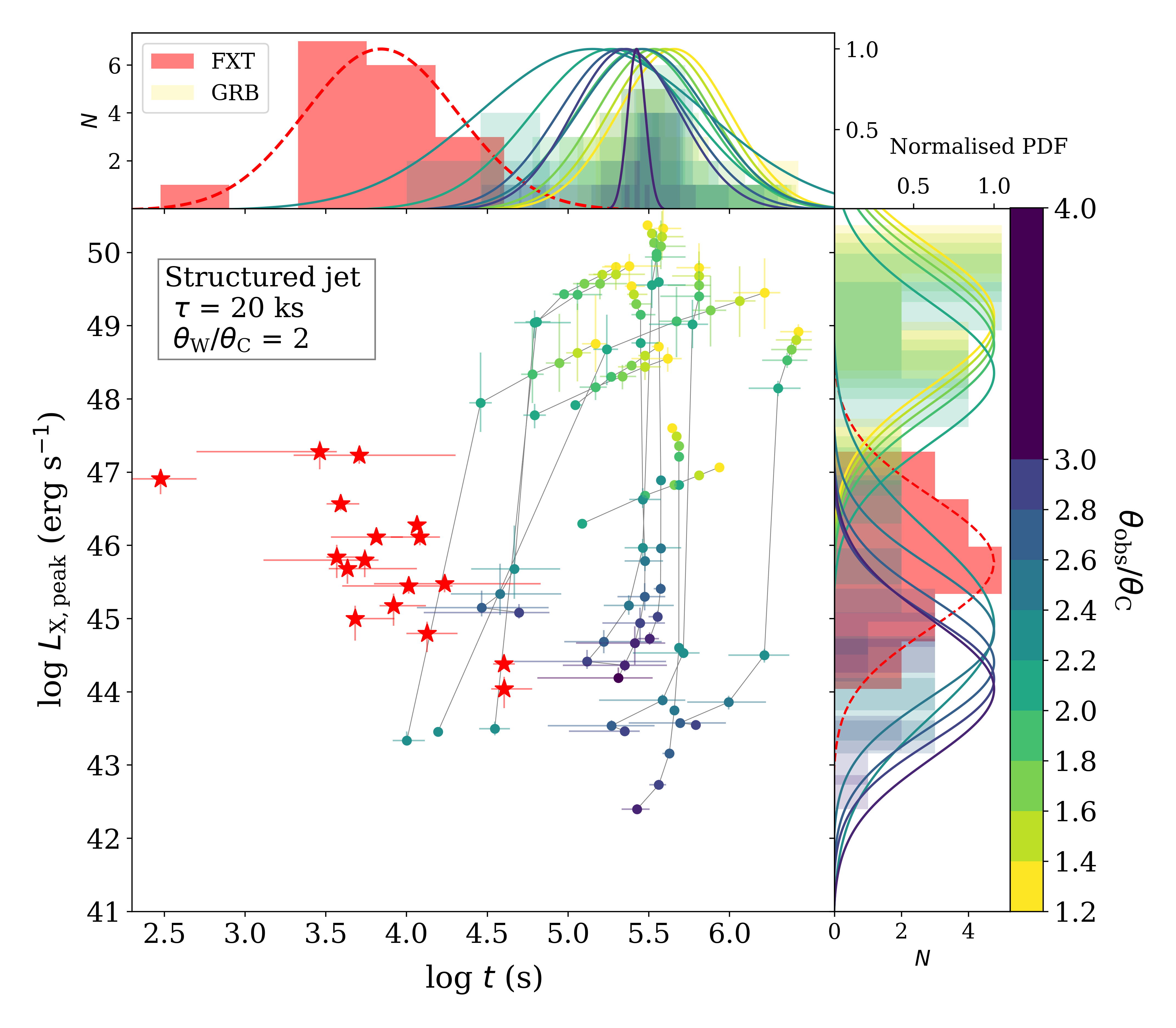}
\caption{Same as the right panel of Fig. 1 but with the isotropic-equivalent kinetic energy, $E_{0}$, of each GRB in our sample decreased (left panel) or increased (right panel) by a factor ten to generate an intrinsically less or more luminous population of GRB afterglows (while keeping fixed all the other afterglow parameters), respectively.}
\label{fig:X-rays-III}
\end{figure*}

\begin{figure}[!ht]
	\includegraphics[width=\columnwidth]{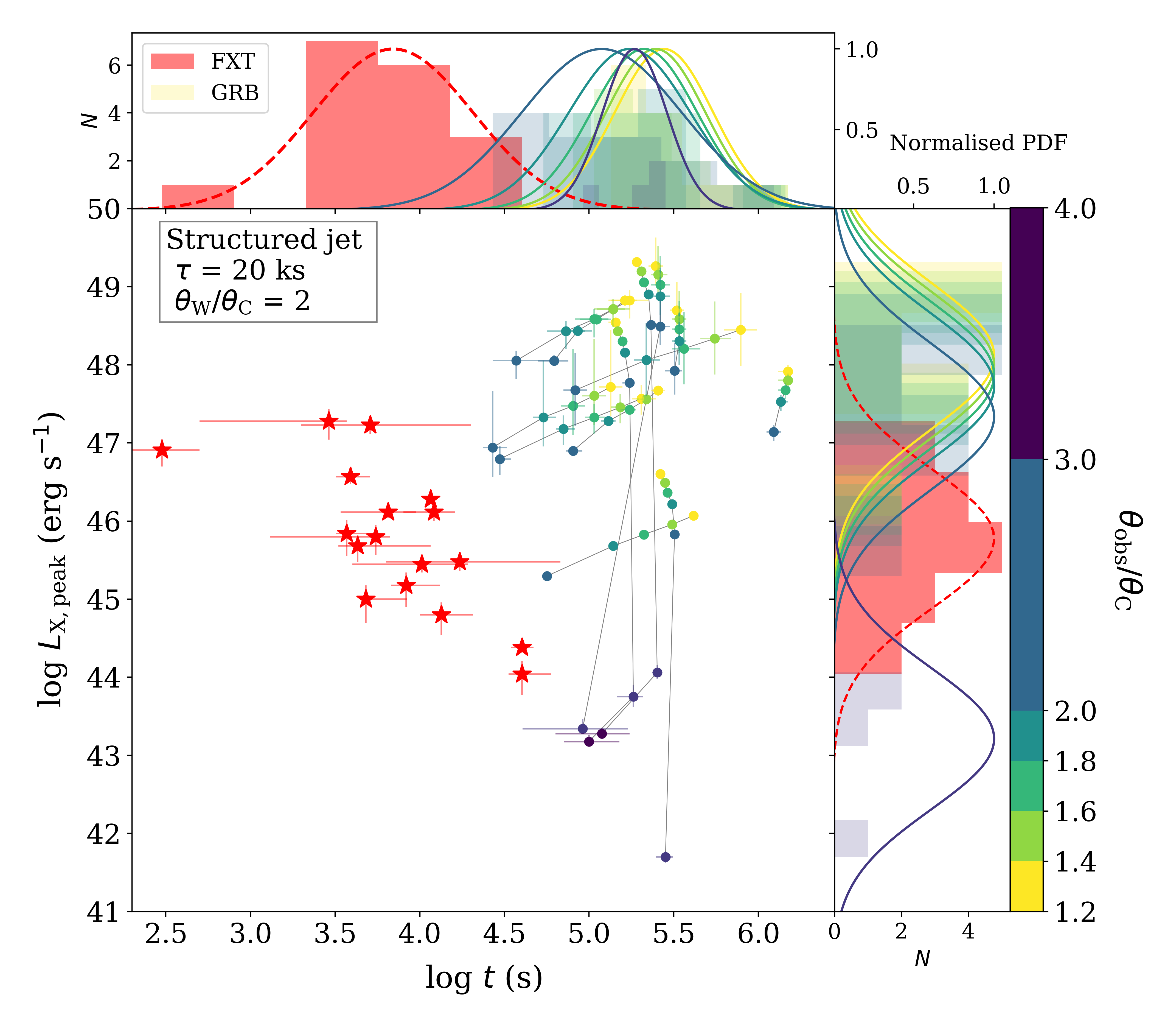}
\caption{Similar figure as the one in the right panel of Fig.~\ref{fig:X-rays} in the main text, but in which the flux limit was reduced by a factor ten (such that it equals the nominal flux limit of \textit{Chandra}).}
\label{fig:X-rays-IV}
\end{figure}

\begin{figure}[!ht]
	\includegraphics[width=\columnwidth]{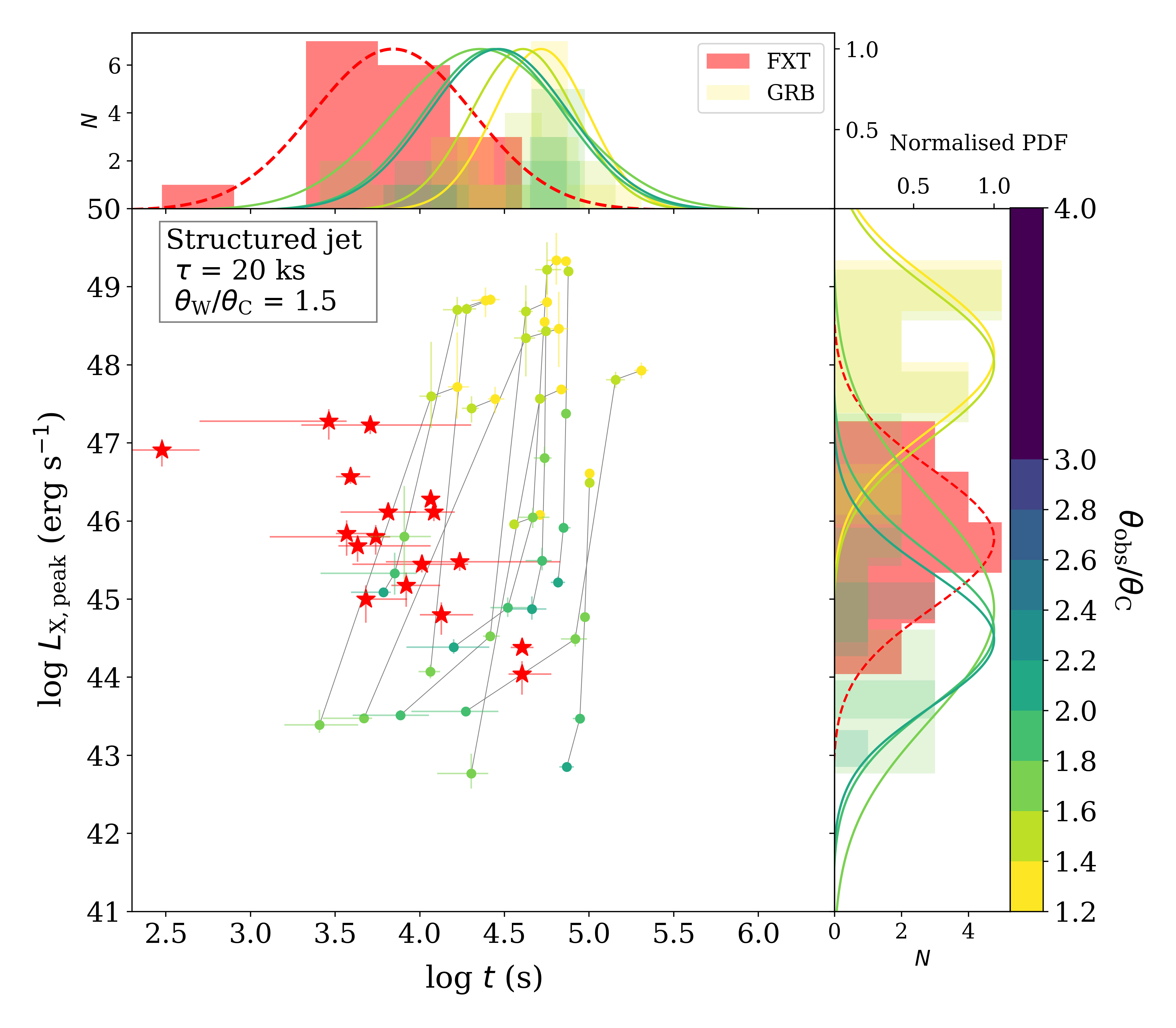}
\caption{Same as the right panel of Fig.~\ref{fig:X-rays} in the main text, here we instead show the results for a structured-jet model with $\theta_{\rm W}$ = 1.5$\theta_{\rm C}$.}
\label{fig:X-rays-II}
\end{figure}

In order to better understand the limitations of the  parameter space we considered, we have varied several parameters, including the inferred isotropic-equivalent kinetic energy $E_{0}$ of the GRBs in our sample, the assumed flux limit of \textit{Chandra}, the truncation angle $\theta_{\rm W}$, and the integration time $\tau$. Fig.~\ref{fig:X-rays-III} and Fig.~\ref{fig:X-rays-IV} show the results derived from changing the isotropic-equivalent kinetic energy $E_{0}$ and the assumed flux limit of \textit{Chandra}: these are discussed in the Discussion section in the main text. Here, we focus on the discussion on the changes in $\theta_{\rm W}$ and $\tau$.

We have performed the same analysis reported in the main text for a structured jet with truncation angle $\theta_{\rm W}$ = 2$\theta_{\rm C}$ (Case I), instead assuming $\theta_{\rm W}$ = 1.5$\theta_{\rm C}$ (Case II). The resulting peak X-ray luminosity and duration of the off-axis afterglows of our selected 13 merger-induced GRBs with a such structured jet are shown in Fig.~\ref{fig:X-rays-II}. For $\theta_{\text{obs}}\leq$ 1.6$\theta_{\rm C}$, the luminosities of the afterglows in Case I and Case II are similar, since the line of sight still lies within the structured wings of the jet in both cases. The luminosity decreases suddenly once $\theta_{\text{obs}}>$ $\theta_{\rm W}$, such that the afterglows are less luminous in the range of viewing angles $\theta_{\text{obs}}\approx (1.6-2)\theta_{\rm C}$ in Case~II compared to Case~I (see Fig.~\ref{fig:X-rays}); for $\theta_{\text{obs}}\gtrsim 2\theta_{\rm C}$ in Case II, the afterglows vanish below the assumed flux limit (we recall here that this corresponds to 10 $\times$ the nominal \textit{Chandra} flux limit). This shows that the luminosity of the off-axis afterglows at different viewing angles strongly depends on the choice of $\theta_{\rm W}$, as expected. 

We then compare the sample of off-axis GRB afterglows with the one of FXTs, as done in the main text. The luminosity distributions of both samples are consistent at a viewing angle $\theta_{\text{obs}} = 1.8\theta_{\rm C}$ ($p = 0.22$), and their duration distributions for $\theta_{\text{obs}} = 2\theta_{\rm C}$ ($p = 0.06$), although we note that at the latter inclination only four out of 13 GRBs in the sample remain detectable. We thus find no viewing angle at which both the luminosity and duration distributions of the FXTs are consistent with the off-axis afterglows simultaneously in Case~II while some were for Case~I (see main text), showing that our results are somewhat sensitive to the choice of $\theta_{\rm W}$. However, considering single events, also in Case II ($\theta_{\rm W}$ = 1.5$\theta_{\rm C}$) we find that the best condition to match the luminosity and durations is to be slightly beyond the truncation angle, here between 1.6 and 2.0 $\theta_{\rm C}$.

Considering the temporal indices of individual FXTs, we recover a shallow segment with temporal index $\alpha_{\text{intermediate}}$ in the off-axis afterglow light curves in Case II that is consistent with the plateau indices of XRT 151121 and XRT 191223 if $\theta_{\text{obs}}= 1.8\theta_{\rm C}$ and of XRT 161125 if $\theta_{\text{obs}}= 1.6, 1.8\theta_{\rm C}$. In the case of XRT 151121 and XRT 191223, the peak X-ray luminosities lie in the bright end of the afterglow luminosity distribution at the corresponding viewing angles. For XRT 161125 this is the case for $\theta_{\text{obs}}= 1.6\theta_{\rm C}$ but not for $\theta_{\text{obs}}= 1.8\theta_{\rm C}$. Also the duration of XRT 161125 is similar only to the shortest-lasting GRB afterglow in the sample that is still detectable at $\theta_{\text{obs}}= 1.6\theta_{\rm C}$. For XRT 151121 and XRT 191223, their durations are 0.1 dex and 0.3 dex shorter than the shortest-lasting GRB afterglow at $\theta_{\text{obs}}= 1.8\theta_{\rm C}$, such that they could only marginally fit into the luminosity-duration phase space of off-axis GRB afterglows at this viewing angle.

We have also studied how the duration changes when choosing a different \textit{Chandra}/ACIS flux limit, namely one corresponding to an integration time of $\tau$ = 10 ks (instead of $\tau$ = 20 ks used in the main text). Averaged over all viewing angles $\theta_{\text{obs}}/\theta_{\rm C}$, the mean afterglow durations are a factor $\approx$ 1.4, 1.6 shorter for an integration time of 10 ks instead of 20 ks (in case of a uniform jet and structured jet, respectively). Still, we find that there is only one viewing angle $\theta_{\text{obs}} = 2.2\theta_{\rm C}$ in the structured-jet case for which both the luminosity distributions ($p$ = 0.10) and duration distributions ($p$ = 0.19) are consistent, which is the same inclination as in the $\tau$ = 20 ks case.

We thus conclude that varying the value of the truncation angle $\theta_{\rm W}$ and the integration time $\tau$ (which determines the \textit{Chandra}/ACIS flux sensitivity limit) may affect the results in detail, but it does not change our overall conclusions, namely that the three FXT observational properties (peak X-ray luminosity, duration and temporal indices) can in individual cases be consistent with merger-driven GRBs when observed slightly beyond the truncation angle $\theta_{\rm W}$ (for Case II with $\theta_{\rm W}$ = 1.5$\theta_{\rm C}$, between 1.6 and 2 $\theta_{\rm C}$, and for Case I with $\theta_{\rm W}$ = 2$\theta_{\rm C}$, between 2.2 and 3$\theta_{\rm C}$).

\newpage
\section{Sample of FXTs and of merger-induced GRBs}
\label{sec:appendix-GRBs}

\subsection{FXTs}
Table~\ref{tab:FXRT} shows the 17 FXTs considered in this research, including their redshift, pre- and post-break temporal indices, peak X-ray luminosity in the 0.3--10 keV band, and duration. A dagger denotes a fiducial redshift; we note that \citet{2022Eappachen} did not find an associated host for XRT 110103, such that the redshift of the nearest member of the galaxy cluster ACO 3581 was used as a fiducial redshift (see \citealt{2015Glennie,2022QuirolaVasquez}), and for XRT 000519 and XRT 161125, the host galaxy association is uncertain.
\begin{table*}
\caption{Properties of the sample of distant FXTs from \citet{2022QuirolaVasquez, 2023QuirolaVasquez}.}
\centering
    \begin{tabular}{cccccc} \hline
    FXT & $z$ & $\alpha_{(1)}$ & $\alpha_{\text{2}}$ & $L_{X,\text{peak}}$ & $T_{90}$ \\
    &  &  &  & (erg s$^{-1}$) & (ks)  \\ \hline
    XRT 000519 &  0.1866$^{\dag}$ & - & - & $1.9\times10^{46}$ & $11.6_{-0.9}^{+1.0}$  \\
    \vspace{0.2cm}
    XRT 030511 & 1.0$^{\dag}$ & 0.2$\pm$0.1 & -1.6$\pm$0.1 & $(1.3\pm 0.2)\times10^{46}$ & $6.5_{-3.1}^{+3.0}$ \\
    \vspace{0.1cm}
    XRT 041230 & 0.61$_{-0.17}^{+0.13}$ & 0.2$\pm$0.1 & -2.0$\pm$0.1 & $(1.1\pm 0.5)\times10^{44}$ & $40.2_{-6.7}^{+19.7}$ \\
    \vspace{0.1cm}
    XRT 080819 & 0.7$_{-0.10}^{+0.04}$ & 0.2$\pm$0.2 & -2.8$\pm$1.9 & $(1.5\pm 0.7)\times10^{45}$ & $8.3_{-1.5}^{+4.9}$  \\
    \vspace{0.1cm}
    XRT 100831 & 1.0$^{\dag}$ & 0.0$\pm$0.1 & -2.4$\pm$0.4 & $(4.8\pm 1.8)\times10^{45}$ & $4.3_{-1.0}^{+7.3}$ \\
    \vspace{0.1cm}
    XRT 110103 & 0.0216$^{\dag}$ & - & - & $(2.4\pm 0.2)\times10^{44}$ & $40.1_{-5.7}^{+6.8}$ \\
    \vspace{0.1cm}
    XRT 110919 & 1.0$^{\dag}$ & 0.0$\pm$0.1 & -1.9$\pm$0.2 & $(3.0\pm 0.7)\times10^{45}$ & $17.2_{-10.9}^{+50.4}$ \\
    \vspace{0.1cm}
    XRT 140327 & 1.0$^{\dag}$ & -0.2$\pm$0.0 & - & $(6.3\pm 2.8)\times10^{44}$ & $13.4_{-3.4}^{+7.3}$ \\
    \vspace{0.1cm}
    XRT 141001 (XT1) & 2.23$_{-1.84}^{+0.98}$ & 0.4 $\pm$0.1 & -1.6$\pm$0.1 & $(1.7\pm 0.4)\times10^{47}$ & $5.1_{-3.1}^{+15.0}$ \\
    \vspace{0.1cm}
    XRT 140507 & 1.0$^{\dag}$ & 0.4$\pm$0.1 & -2.9$\pm$1.2 & $(1.0\pm 0.5)\times10^{45}$ & $4.8_{-5.5}^{+3.6}$ \\
    \vspace{0.1cm}
    XRT 150322 (XT2) & 0.738 & -0.09$\pm$0.1 & -2.0$\pm$0.3 & $(2.8\pm 0.6)\times10^{45}$ & $10.3_{-6.3}^{+9.0}$ \\
    \vspace{0.1cm}
    XRT 151121 & 1.0$^{\dag}$ & -0.3$\pm$0.1 & - & $(6.3\pm 2.6)\times10^{45}$ & $5.5_{-4.2}^{+1.2}$  \\
    \vspace{0.1cm}
    XRT 161125 & 0.35$^{\dag}$ & -0.5$\pm$0.1 & - & $(1.9\pm 0.8)\times10^{47}$ & $2.9_{-2.4}^{+0.8}$  \\
    \vspace{0.1cm}
    XRT 170901 & 1.44 & -0.1$\pm$0.1 & -1.9$\pm$0.5 & $(3.7\pm 0.7)\times10^{46}$ & $3.9_{-0.7}^{+1.2}$  \\
    \vspace{0.1cm}
    XRT 191127 & 1.0$^{\dag}$ & 1.0$\pm$0.2 & -2.5$\pm$0.3 & $(8.1\pm 3.1)\times10^{46}$ & $0.3 \pm 0.2$ \\
    \vspace{0.1cm}
    XRT 191223 & 0.85 & -0.4$\pm$0.1 & - & $(6.9\pm 3.3)\times10^{45}$ & $3.7_{-0.5}^{+2.4}$ \\
    \vspace{0.1cm}
    XRT 210423 & 1.5105 & -0.2$\pm$0.1 & -3.8$\pm$1.2 & $(1.3\pm 0.3)\times10^{46}$ & $12.1_{-4.1}^{+4.0}$  \\
    
    \end{tabular}
\tablefoot{Content of columns: (1) FXT, (2) redshift, (3) pre-break temporal index (or the single temporal index in case the light curve is best-fit by a simple power law), (4) post-break temporal index, (5) X-ray peak luminosity, (6) duration over which central 90\% of \textit{Chandra} 0.5--7 keV photon counts are detected. Values were taken from \citet{2022QuirolaVasquez, 2023QuirolaVasquez}. The sign of the temporal indices follows from the notation $F\propto t^{\alpha}$.
\\
$^{\dag}$ Fiducial redshift.}
\label{tab:FXRT}
\end{table*}

\subsection{Merger-Induced GRBs}
\begin{table*}
\caption{Final Sample of 13 merger-induced GRBs analysed in this work.} 
    \centering
    \makebox[1 \textwidth][c]{
    \resizebox{1. \textwidth}{!}{
    
    \begin{tabular}{cccccc} \hline
    GRB &  $z$ & $\Gamma_{X}$ & $T_{90}$ (EE) (s) & References & References optical data\\ \hline
    050709 & 0.160 $\pm$ 0.001 & 2.24$\pm$0.35$^{\dag}$ &0.07$\pm$0.01 (130$\pm$7) & [1], [2] & [1], [24], [25], [26], [27], [28], [29] \\
    \vspace{0.2cm}
    050724 & 0.2582 $\pm$ 0.0003 & 1.80$^{+0.05}_{-0.04}$ & 3.0$\pm$0.1 (152.4$\pm$9.2) & [3], [4] & [3], [4], [29], [30] \\
    \vspace{0.1cm}
    051227 & 1 assumed & 1.85$^{+0.12}_{-0.09}$ & 8.0$\pm$0.2 ($\sim$100) & [5], [6] & [5], [31], [32], [33], [34], \\
    &  &  &  &  & [35], [36], [37], [38] \\
    \vspace{0.1cm}
    061006 & 0.436 $\pm$ 0.002 & 2.03$^{+0.11}_{-0.08}$ & $\sim$0.5 (130$\pm$10) &[5], [7] & [5], [29], [37], [39], [40], [41] \\
    \vspace{0.1cm}
    061201 & 0.111 uncertain & 1.87$^{+0.19}_{-0.14}$ & 0.8$\pm$0.1 & [8] & [8], [42], [43], [44] \\
    \vspace{0.1cm}
    070714B & 0.9225 $\pm$ 0.0001 & 1.66$^{+0.11}_{-0.08}$ & $\sim$3 ($\sim$100) & [9], [10] & [45], [46], [47], [48], [49], [50] \\
    \vspace{0.1cm}
    070724A & 0.457 $\pm$ 0.0007 & 1.78$^{+0.13}_{-0.10}$ & 0.4$\pm$0.04 & [11] & [11], [29], [51], [52], [53] \\
    \vspace{0.1cm}
    071227 & 0.381 & 1.64$^{+0.14}_{-0.11}$ & 1.8$\pm$0.4 ($\gtrsim$300) & [5], [12] & [5], [29], [54], [55], [56], [57], [58], [59] \\
    \vspace{0.1cm}
    090510 & 0.903 $\pm$ 0.001 & 2.05$^{+0.12}_{-0.09}$ & 0.3$\pm$0.1 ($\sim$100) & [13], [14], [15] & [29], [59], [60], [61], [62] \\
    \vspace{0.1cm}
    110112A & 0.5 assumed & 2.04$^{+0.09}_{-0.07}$ & 0.5$\pm$0.1 & [16], [17] & [63], [64], [65], [66], [67], [68]\\
    \vspace{0.1cm}
    111020A & 0.5 assumed & 1.70$^{+0.10}_{-0.07}$ & 0.40$\pm$0.09 & [18], [19] &  \\
    \vspace{0.1cm}
    111117A & 2.211 $\pm$ 0.001 & 1.63$^{+0.09}_{-0.07}$ & 0.47$\pm$0.09 &  [20], [21] & \\
    \vspace{0.1cm}
    160821B & 0.1619 $\pm$ 0.0002 & 1.88$\pm$0.24$^{\dag\dag}$ & 0.48$\pm$0.07 & [22], [23] & [69], [70], [71], [72], [73] \\
    \end{tabular}
    }}
\tablefoot{Columns: (1) name of GRB, (2) redshift, (3) X-ray photon index (averaged over times when detections were included in the fitting), (4) duration $T_{90}$ over which central 90\% of the total observed 15--150 keV photon counts are detected (in parentheses: duration including the duration of the extended emission (EE) if $T_{90}>$ 2s; see \citealt{2011Kann}), (5) references for columns (2)--(4), (6) references for the optical data.
\\
$^{\dag}$ Photon index taken from \citealt{2005Fox}.
\\
$^{\dag\dag}$ Photon index taken from \citealt{2019Troja}.
\\
References: [1] \citealt{2005Fox}, [2] \citealt{2009Nysewander}, [3] \citealt{2007Malesani}, [4] \citealt{2005Berger}, [5] \citealt{2009DAvanzo}, [6] \citealt{2005Hullinger}, [7] \citealt{2006Krimm}, [8] \citealt{2007Stratta}, [9] \citealt{2017Gao}, [10] \citealt{2007Barbier}, [11] 
\citealt{2010Kocevski}, [12] \citealt{2007Sato}, [13] \citealt{2010McBreen}, [14] \citealt{2010Ackermann}, [15] \citealt{2009Grupe}, [16] \citealt{2013Fong}, [17] \citealt{2011Barthelmy}, [18] \citealt{2012Fong}, [19] \citealt{2011Sakamoto}, [20] \citealt{2018Selsing}, [21] \citealt{2011bSakamoto}, [22] \citealt{2022Fong}, [23] \citealt{2016Palmer}, [24] \citealt{2005Hjorth}, [25] \citealt{2006Covino}, [26] \citealt{2005Morgan}, [27] \citealt{2005Haislip}, [28] \citealt{2005Monard}, [29] \citealt{2010Leibler}, [30] \citealt{2006Gorosabel}, [31] \citealt{2005Roming}, [32] \citealt{2006Kodaka}, [33] \citealt{2005Yanagisawa}, [34] \citealt{2005Cenko}, [35] \citealt{2005Garimella}, [36] \citealt{2005Sonoda}, [37] \citealt{2007Berger}, [38] \citealt{2005Halpern}, [39] \citealt{2006Schady}, [40] \citealt{2006Mundell}, [41] \citealt{2010Fong}, [42] \citealt{2006Holland}, [43] \citealt{2006Marshall}, [44] \citealt{2010Berger}, [45] \citealt{2007Landsman}, [46] \citealt{2007Weaver}, [47] \citealt{2008Melandri}, [48] \citealt{2009Graham}, [49] \citealt{2007Perley}, [50] \citealt{2011Kann}, [51] \citealt{2007Ziaeepour}, [52] \citealt{2009Berger}, [53] \citealt{2007Cenko}, [54] \citealt{2007Cucchiara} , [55] \citealt{2007Sakamoto}, [56] \citealt{2019Eyles}, [57] \citealt{2007BMR}, [58] \citealt{2007DAvanzo}, [59] \citealt{2012NicuesaGuelbenzu}, [60] \citealt{2010DePasquale}, [61] \citealt{2009Olofsson}, [62] \citealt{2010McBreen}, [63] \citealt{2011Breeveld}, [64] \citealt{2011Gorbovskoy}, [65] \citealt{2011Xin}, [66] \citealt{2011Nakajima}, [67] \citealt{2011Grankin}, [68] \citealt{2013Fong}, [69] \citealt{2016Breeveld}, [70] \citealt{2019Troja}, [71] \citealt{2016Rebolo}, [72] \citealt{2019Lamb}, [73] \citealt{2018Jin}.
}
\label{tab:sGRB}
\end{table*}

Table~\ref{tab:sGRB} provides the names, redshifts, photon indices, and durations of the final sample of 13 merger-induced GRBs used in this research. Here, we describe the data selection of each burst.
\\ \\
\textit{GRB 050709}. X-ray data (\textit{HETE}, \textit{Chandra} and \textit{Swift} flux densities at 5 keV) and radio upper limits are taken from table 1 by \citet{2005Fox}. They attribute the early HETE detection at 100 s and \textit{Chandra} detection at 16 days to flaring, so we treat these detections as upper limits. The 5 keV flux densities were converted to 10 keV flux densities assuming the best-fitting photon index $\Gamma = 2.24 \pm 0.35$ found by \citet{2005Fox}. 
\\ \\
\textit{GRB 050724}. The X-ray light curve of GRB 050724 contains a large flare identified at $1.20\times10^{4}-1.10\times10^{5}$ s in the \textit{Swift}/XRT Catalogue; detections belonging to this flare are treated as upper limits. Two radio detections from \citet{2005Berger} and five additional \textit{Chandra} detections by \citet{2006Grupe} are included, although only the last detection ($> 10^{6}$ s) connects to the assumed afterglow light curve such that the rest is treated as upper limits. The radio data are not necessarily affected by the flaring in the X-ray band according to \citet{2007Malesani}, although the optical data could be related to the flare, such that optical detections before $1.10\times10^{5}$ s are treated as upper limits as well. A photon index of $\Gamma$ = 1.85 was used to convert the 0.3--10 keV flux to the 10 keV flux density \citep{2006Grupe}.
\\ \\
\textit{GRB 051227}. The X-ray light curve of GRB 051227 contains a tail of the prompt emission \citep{2009DAvanzo} and superposed flares at early times, detections of which are treated as upper limits here, followed by the standard forward shock emission \citep{2009Dado}. GRB 051227 has optical detections in addition.
\\ \\
\textit{GRB 061006}. The X-ray light curve of GRB 061006 decays as a simple power-law, and there are two optical detections for this burst. However, at later times, the X-ray and optical light curve behaviour becomes inconsistent; the late time optical data of GRB 061006 are likely part of the merger-nova emission powered by a magnetar according to \citet{2017Gao} (and also \citealt{2009DAvanzo} attribute the steep optical decay to an additional optical component). For this reason, the optical detections are treated as upper limits. 
\\ \\
\textit{GRB 061201}. The X-ray light curve of GRB 061201 has a break at $t_{b}$ = 2.39$^{+0.58}_{-0.52}$ ks, and the pre- and post-break temporal decay indices are $\alpha_{1} = -0.54 \pm  0.08$ and $\alpha_{2} = -1.90 \pm 0.15$, respectively \citep{2007Stratta}. The pre- and post-break phases could correspond to a plateau phase and normal decay phase; alternatively, the steepening can be interpreted as a jet break, although it is not certain whether this break is achromatic \citep{2007Stratta}. We assume it is and include all optical detections. Since $|\alpha_{1}| <$ 0.8 in the X-ray light curve, we assume this phase to be a plateau, and thus treat these X-ray detections as upper limits. 
\\ \\
\textit{GRB 070714B}. The X-ray light curve of GRB 070714B first decays with $\alpha = -2.49\pm 0.18$ until a plateau starts at $\sim$ 400 s; the slope of the plateau is then $\alpha = -0.60 \pm 0.29$, and the plateau ends at $\sim$ 1000 s after which the light curve decays as $\alpha  = -1.73 \pm 0.11$ \citep{2017Gao}. \citet{2017Gao} suggest that the late-time optical detections belong to emission from a magnetar-powered merger-nova. For this reason, the two optical detections at late times were treated as upper limits, and so were the X-ray detections that are part of the initial steep decay and plateau. A radio upper limit from \citet{2015Fong} was included in the fit.
\\ \\
\textit{GRB 070724A}. The early X-ray light curve of GRB 070724A consists of a steep decay, which is treated as upper limits; the data beyond $\sim$ 300 s are consistent with a simple power-law decay \citep{2010Kocevski}, with $\alpha = -1.37 \pm 0.03$ . There are three optical detections, and a radio upper limit was included (taken from \citealt{2015Fong}).
\\ \\
\textit{GRB 071227}. The X-ray light curve of GRB 071227 contains a steep decay with a break at 389 $\pm$ 65 s, followed by a decay with index $\alpha = -1.1 \pm 0.2$  \citep{2019Eyles}. The steep decay prior to 389 s was treated as upper limits. There are six optical detections, and no radio data.
\\ \\
\textit{GRB 090510} has an initial rise in the optical light curve, while the X-ray light curve contains a plateau \citep{2009Grupe, 2010DePasquale, 2012bNicuesaGuelbenzu}. Since the X-ray plateau cannot result from a simple forward shock model \citep{2010DePasquale}, we treat detections that are part of it as upper limits.
\\ \\
\textit{GRB 110112A} has a single optical detection, and the X-ray light curve follows a simple power-law decay without any signs of spectral evolution. A radio upper limit was taken from \citet{2015Fong}.
\\ \\
\textit{GRB 111020A}. For this burst, X-ray data were taken from \citet{2012Fong}, as these include \textit{Swift}, XMM and \textit{Chandra} detections and one \textit{Chandra} upper limit. They also includes one radio upper limit. The optical data consist of upper limits only.
\\ \\
\textit{GRB 111117A} has no optical detections, and the X-ray light curve decays as a simple power law. For this burst, a radio upper limit and additional \textit{Chandra} detection were taken from \citet{2015Fong} and included in the fitting.
\\ \\
\textit{GRB 160821B}. For this burst, X-ray data are taken from table 1 by \citet{2019Troja}, which contains \textit{Swift} detections, two XMM detections, a \textit{Swift} upper limit, a radio detection, and radio upper limits. These are reported in units of flux density ($\mu$Jy), and corrected for Galactic absorption. The single radio detection is likely due to reverse shock emission, and was thus excluded from broadband modelling by \citet{2019Troja}. Here, it is  treated as an upper limit as well. \citet{2019Troja} include optical data up to 0.08 days in their fit; the rest of the optical data ($\gtrsim$ 1 day) are part of the kilonova emission and are treated as upper limits. In our case, the contribution from the kilonova has already been subtracted from the optical data of GRB 160821B, such that we include data beyond 1 day in the fitting.

\section{Fitting results}
\label{sec:appendix-fitting}

In order to model the GRB afterglows, we converted the unabsorbed 0.3--10 keV fluxes to flux densities at 10 keV as
\begin{equation}
    F_{\text{0.3--10 keV}} = \frac{(\text{10 keV})^{2-\Gamma_{X}}-(\text{0.3 keV})^{2-\Gamma_{X}}}{h(2-\Gamma_{X})(\text{10 keV})^{1-\Gamma_{X}}}\times F_{\text{10 keV}},
\end{equation}
where a constant $\Gamma_{X}$ was assumed. When converting the flux densities back to fluxes and, subsequently, the luminosities in the 0.3--10 keV band, we used that
\begin{equation}
    L_{\text{0.3--10 keV}} = 4\pi d_{L}^{2}(z)F_{\text{0.3--10 keV}}(1+z)^{\Gamma_{X}-2} 
\end{equation}
where the factor $(1+z)^{\Gamma_{X}-2}$ is the $K$-correction factor (e.g. \citealt{2013Dainotti}).
\\ \\
In our MCMC parameter estimation, the log-likelihood function ln ($\mathcal{L}$) is that of a variable Gaussian, as formulated by \citep{2004Barlow}. It is expressed in terms of the measured flux density $F_{\nu}$, its associated uncertainty $\sigma$ (which may be asymmetric), and model flux density $F_{\nu}(t;\theta)$ as
\begin{equation}
    \text{ln }(\mathcal{L}) = 
    -\frac{1}{2}\sum_{i}\left[\left(\frac{F_{\nu,i}-F_{\nu}(t_{i};\theta)}{
    \sigma_{i} + \sigma'_{i}(F_{\nu}(t_{i};\theta) - F_{\nu,i})
    }\right)^{2}\right] 
\end{equation}
where $\sigma_{i} = (2\sigma_{i,+}\sigma_{i,-})/(\sigma_{i,+}+\sigma_{i,-})$ and $\sigma_{i}' = (\sigma_{i,+}-\sigma_{i,-})(\sigma_{i,+}+\sigma_{i,-})$ (the positive and negative uncertainties are denoted by $\sigma_{i,+}$ and $\sigma_{i,-}$ respectively). In the case of upper limits, we set
\begin{equation}
    \text{ln }(\mathcal{L}_{i}) = 
    \begin{cases}
    0 & F_{\nu}(t_{i};\theta) < F_{\nu,i} \\
    -\infty & F_{\nu}(t_{i};\theta) \geq F_{\nu,i}.
    \end{cases}
\end{equation}

We discuss the fitting results of the GRB sample in a general sense. Table~\ref{tab:fits} displays the best-fitting model parameters for each GRB afterglow obtained through the broadband light curve modelling with \texttt{afterglowpy} + MCMC parameter estimation for the uniform-jet models. For each jet parameter, we report the median value plus 16th and 84th percentile uncertainties. To assess the goodness of fit, we use the statistic -2 ln ($\mathcal{L}$) and apply it to the set of parameters $\Theta$ that maximises the likelihood (the \textit{maximum a posteriori}, or MAP). Figure~\ref{fig:fits} shows the fit to the multi-wavelength light curve of each GRB afterglow. 
\\
Median values of the isotropic-equivalent kinetic energy of the sample of merger-induced GRBs are in the range $E_{0}\sim 10^{50-53}$ erg, with a sample mean of $\sim 6\times 10^{51}$ erg, and the mean circum-burst density is $\sim 2\times 10^{-3}$ cm$^{-3}$. These are consistent with typical parameter values expected for merger-induced GRBs ($E_{0}\sim 10^{49-53}$ erg and median of (2-4)$\times10^{51}$ erg; median density $n_{0}\approx 10^{-3}$ cm$^{-3}$ \citep{2015Fong, 2022RoucoEscorial}. The overall low median values of the circum-burst density $n_{0}$ are consistent with merger-induced GRBs favouring low-density environments; GRB 050709, GRB 061201, GRB 070714B, GRB 071227, GRB 090510, GRB 111020A, GRB 111117A and GRB 160821B are all found in the outskirts of their presumed hosts, while GRB 110112A lacks an associated host galaxy \citep{2005Fox, 2007Stratta, 2010Kocevski, 2019Eyles, 2011Panaitescu, 2012Fong, 2012Margutti, 2019Lamb}. GRB 050724, GRB 051227, and GRB 070724A have relatively high circum-burst densities (compared to the other GRBs in this sample), which is not at odds with the low offsets to their hosts \citep{2005Berger, 2009DAvanzo} or large extinction that might indicate a dusty environment \citep{2009Berger}. Only GRB 061006 has a small host offset \citep{2009DAvanzo} combined with a low circum-burst density, but the latter is almost unconstrained.
\\
For the two bursts that were previously reported to have an observed jet break in their light curves, GRB 061201 and GRB 111020A \citep{2018Jin}, we infer a narrow core angle of $\theta_{\rm C}\lesssim 2^{\circ}$. For GRB 050709, most jet parameters are not well constrained, including $\theta_{\rm C}$; the latter is not surprising, as the lack of a jet break may only provide a lower limit on the opening angle \citep{2018Jin}. For GRB 160821B, various studies also report a jet break at 0.35 to 3.5 days \citep{2018Jin, 2019Troja, 2019Lamb}, depending on whether the two late-time \textit{XMM-Newton} detections are assumed to be part of the post-jet-break decay \citep{2019Troja} or of a refreshed shock at $\approx$ 1 day \citep{2019Lamb}. In our case, the model does not capture a jet break in the observed data (see Fig.~\ref{fig:fits}), likely due to the sparsity of the X-ray detections beyond $\approx$ 1 day and possibly due to contamination of a refreshed shock in both the X-ray and optical bands at this time. This results in a poorly constrained core angle.
\\
A few bursts, such as GRB 051227 and GRB 090510, have well-constrained jet parameters and acceptable fits (considering the reduced -2 ln ($\mathcal{L}$) statistic; see Table~\ref{tab:fits}). This is not the case for many other bursts in the sample, and could be partly due to distinct detections that the model does not capture. For example, the model fits the data of GRB 050724 well at first glance (Fig.~\ref{fig:fits}), except for the late-time \textit{Chandra} detection which the model underestimates by two orders of magnitudes. The optical decay beyond $\sim$ 1 day seems to be interpreted as the post-jet break decay, which follows the interpretation by \citet{2005Berger}. On the other hand, \citet{2007Malesani} find that the optical data are likely related to the X-ray flare, and confirm that the break at $\sim$ 1 day is unlikely to be a jet break. We ran the model again while treating all optical detections as part of the flare and thus as upper limits, but this model did not improve the fit (stat/d.o.f. = 3.22). For GRB 110112A, the single optical detection is not captured by the model, and there is no spectral evolution (Fig.~\ref{fig:fits}) or flaring in the X-ray light curve that hints at a flare in the optical light curve. A late onset of the afterglow seems unlikely considering the steep decay in X-rays. Finally, although the jet parameters of GRB 061201 are well constrained and consistent with results of other authors \citep{2007Stratta, 2015Fong}, the goodness of fit is not acceptable, which might be attributed to flaring or variability in the optical and X-ray light curves. In selecting the data of the GRBs in our sample, we have made attempts to remove any detections that do not belong to the forward shock of the afterglow and thus cannot be modelled with \texttt{afterglowpy}. The often poor reduced -2 ln ($\mathcal{L}$) statistic of the fits suggests not all detections are supported by the model, and possibly originate from another emission source still.

\begin{figure*}
    \includegraphics[width=\columnwidth]{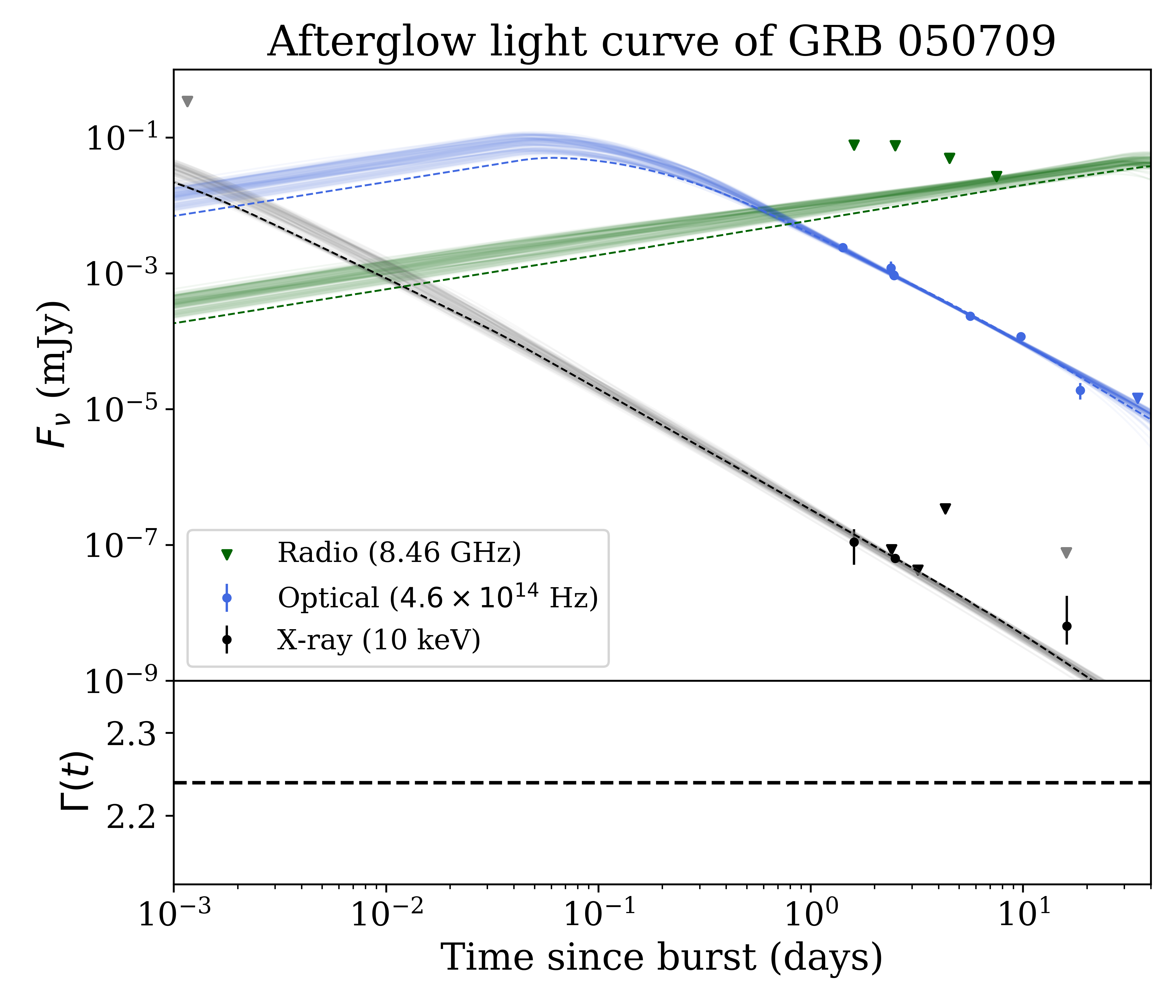}
    \includegraphics[width=\columnwidth]{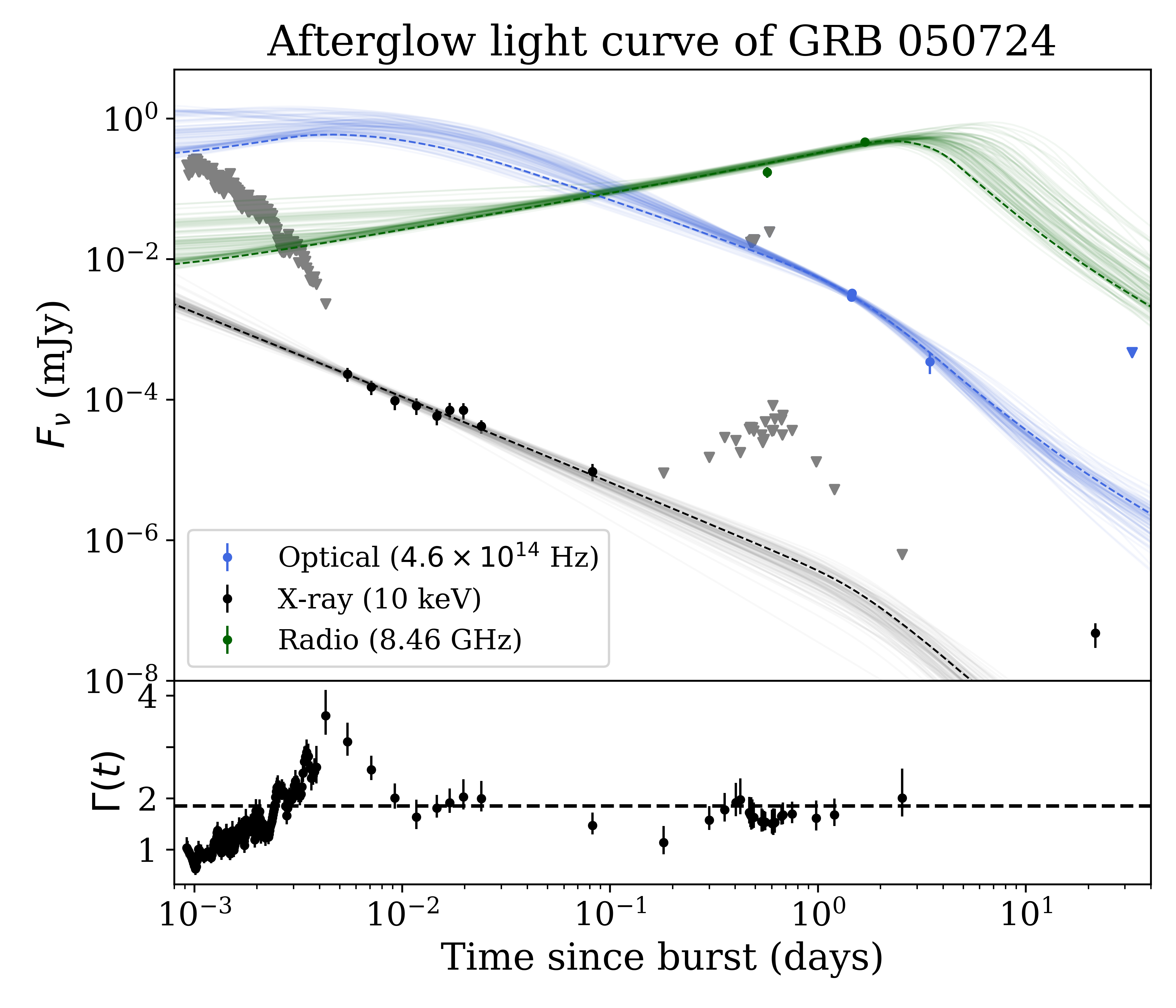}
    \includegraphics[width=\columnwidth]{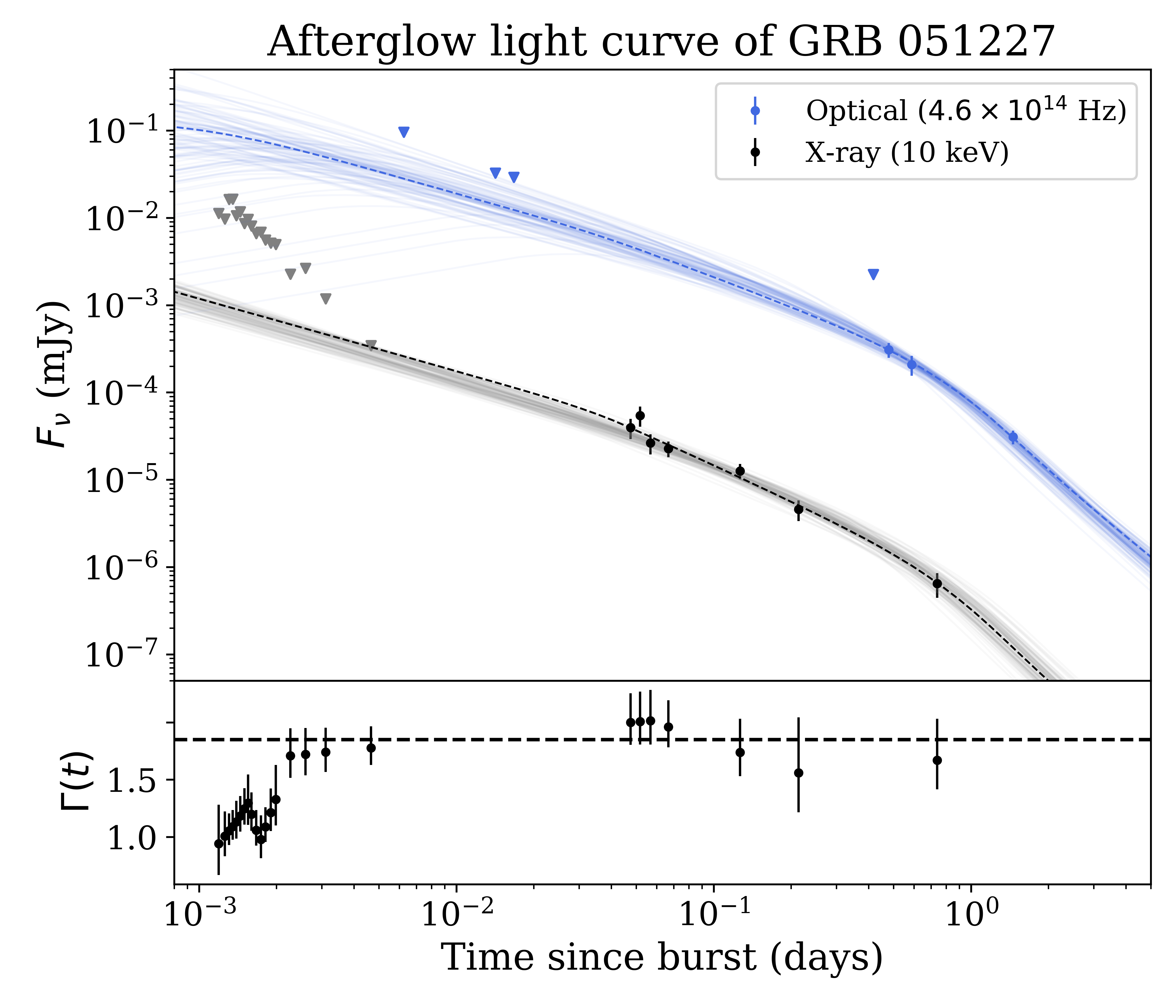}
    \includegraphics[width=\columnwidth]{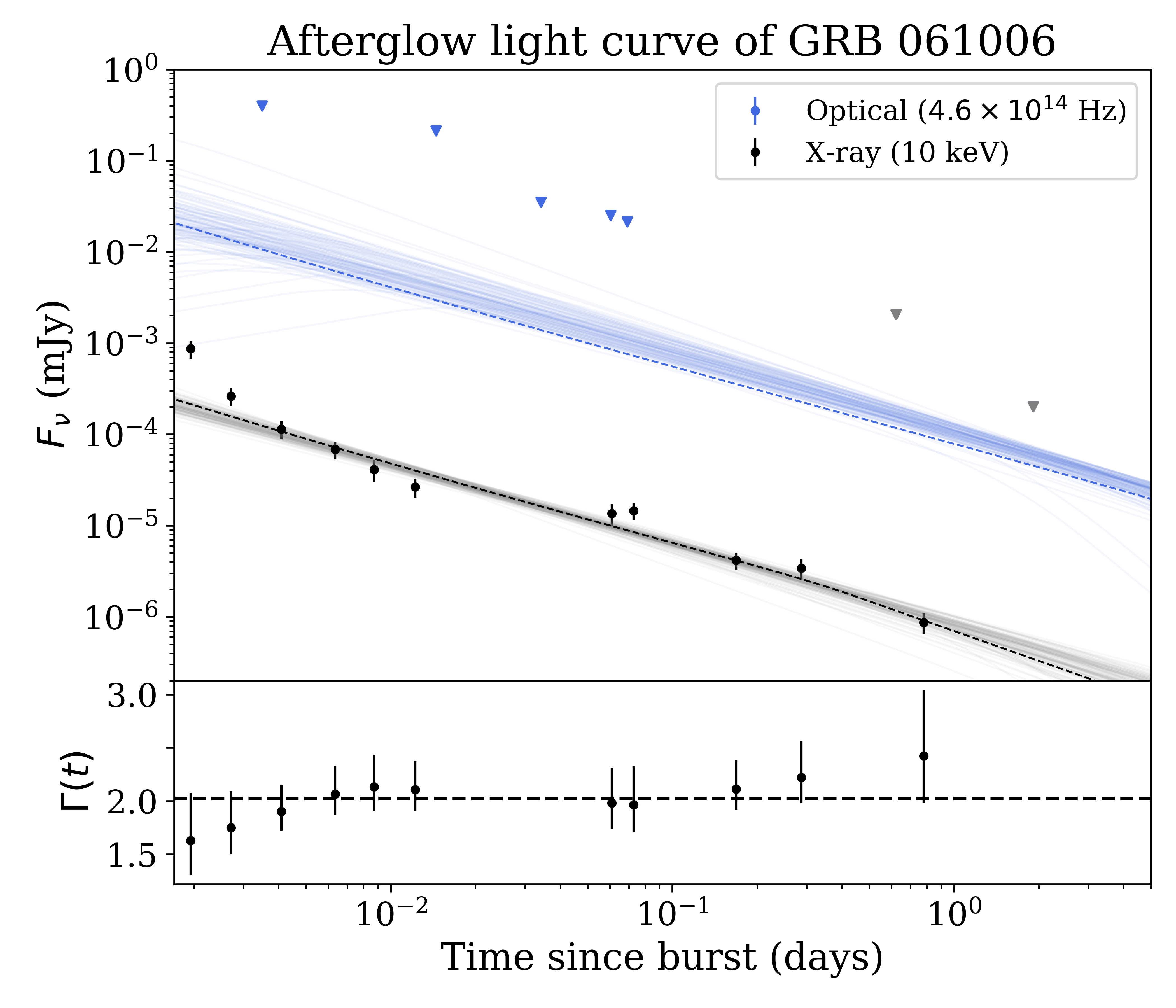}
    \caption{Fits to the broadband light curves of the 13 merger-induced GRB afterglows in the final sample. The upper panel shows the flux density as a function of time since the GRB trigger. Black, blue and green points represent X-ray, optical, and radio detections, respectively; coloured triangles denote true upper limits, while grey triangles represent detections that were treated as upper limits in the fitting procedure as they were deemed to be caused by another emission process than the standard forward shock afterglow emission. Dashed lines mark the best fit (\textit{maximum a posteriori}) to the data, accompanied by a hundred samples randomly drawn from the posterior flux density distribution within the 16th -- 84th percentile range to visualise the uncertainty region. The bottom panel shows the evolution of the photon index of the X-ray spectrum, $\Gamma_{X}(t)$, also taken from the \textit{Swift} Light Curve Repository. For GRB 050709 and GRB 160821B, all X-ray data were taken from other authors rather than from the \textit{Swift} Light Curve Repository, such that the evolution of the photon index is not shown.}
\end{figure*}
\begin{figure*}\ContinuedFloat
    \includegraphics[width=\columnwidth]{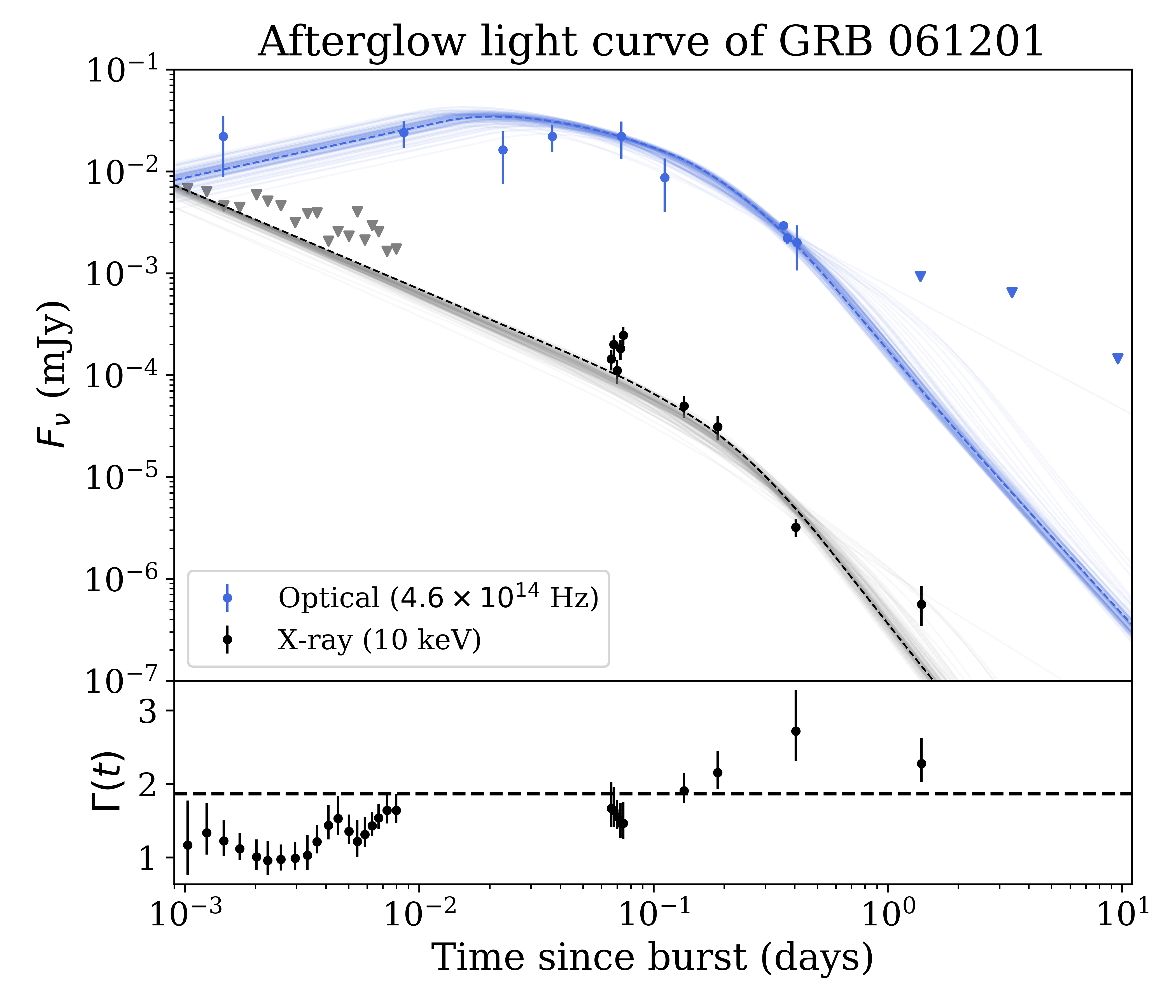}
    \includegraphics[width=\columnwidth]{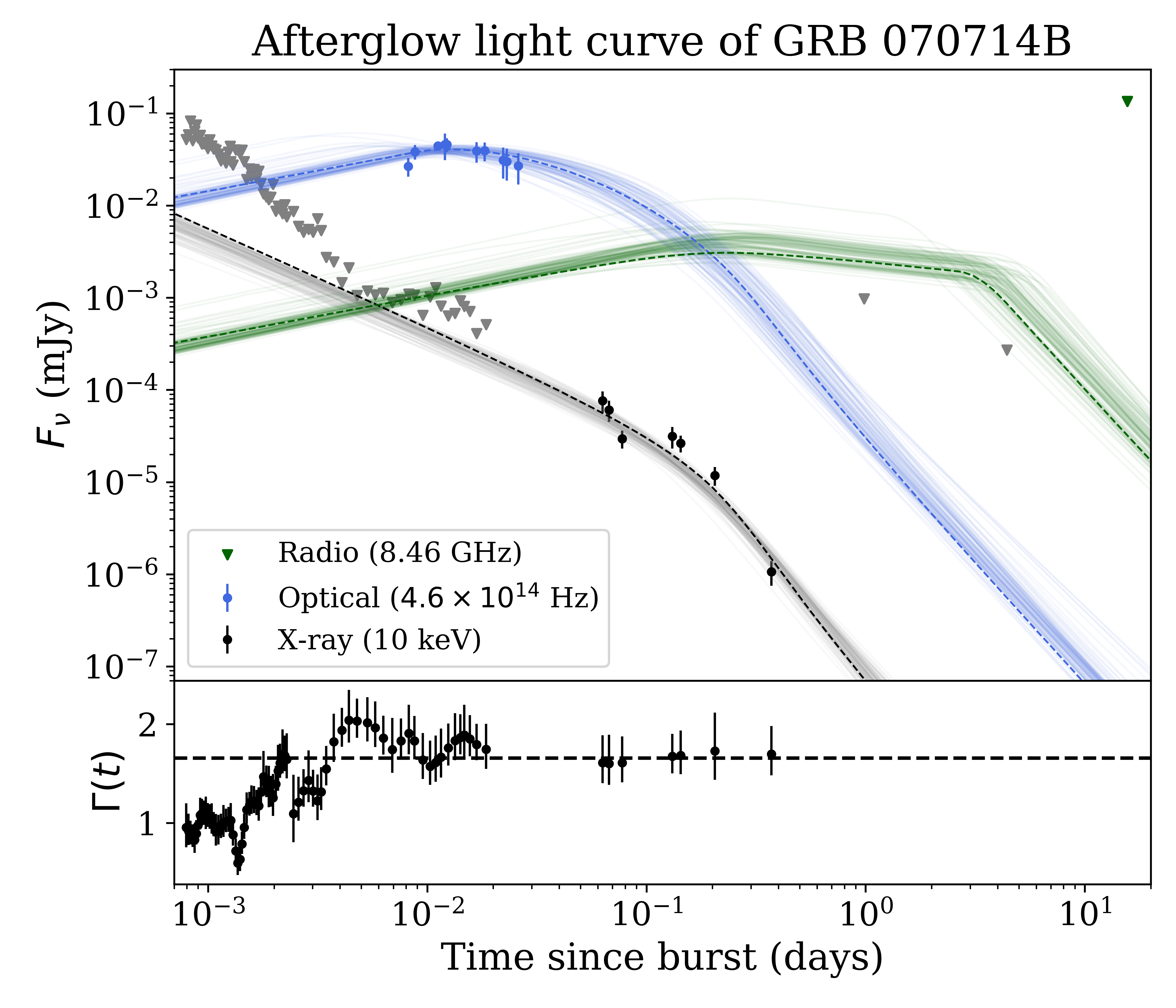}
    \includegraphics[width=\columnwidth]{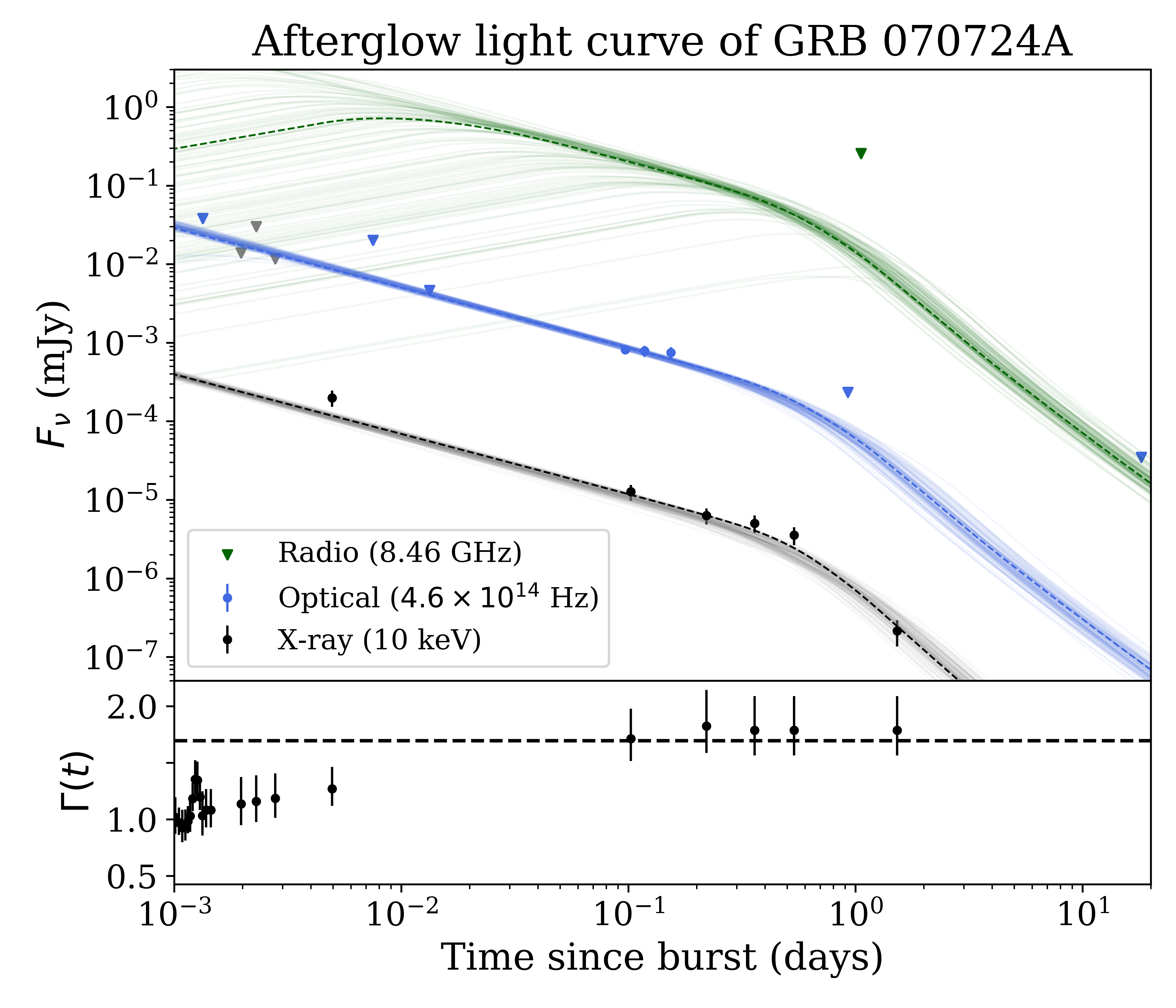}
    \includegraphics[width=\columnwidth]{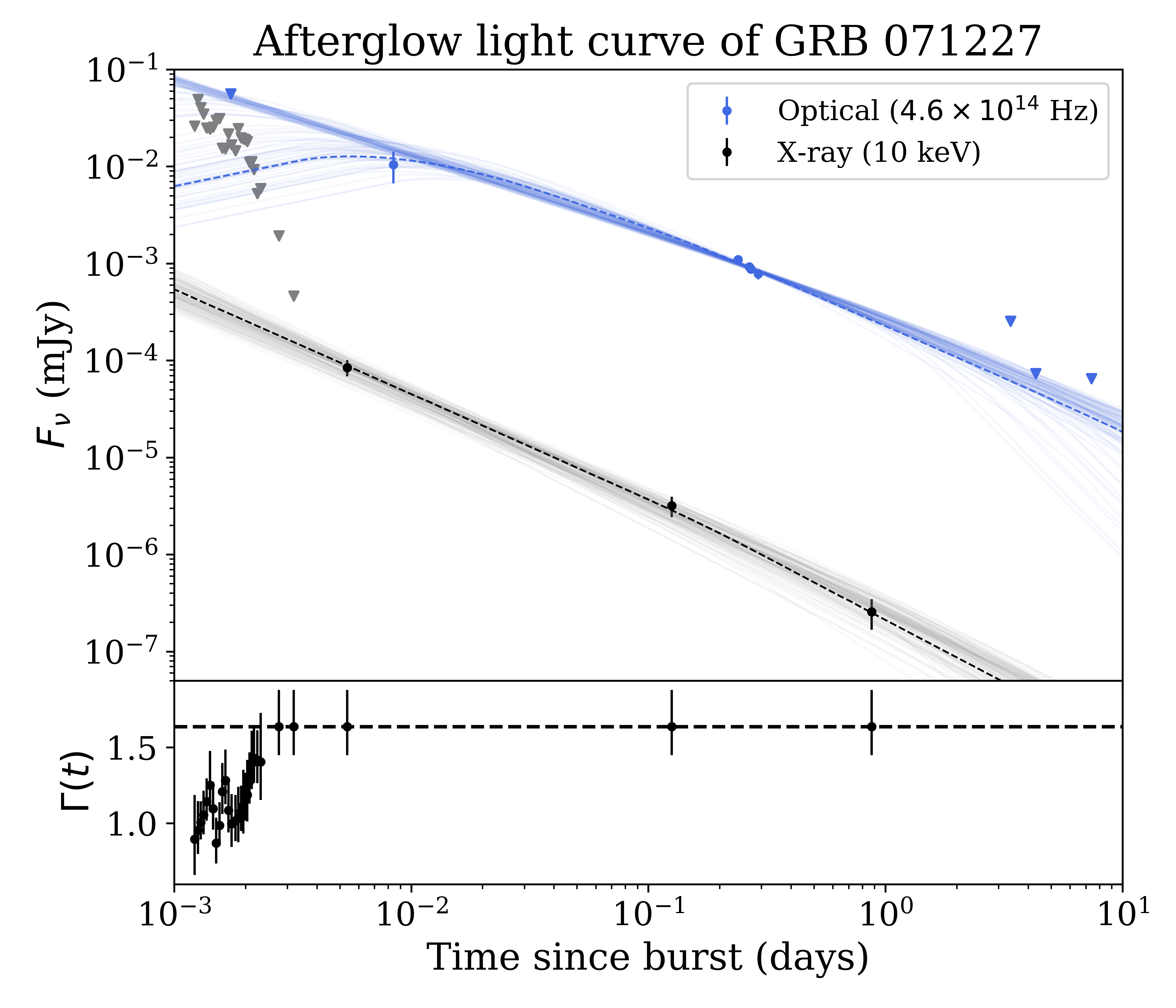}
    \includegraphics[width=\columnwidth]{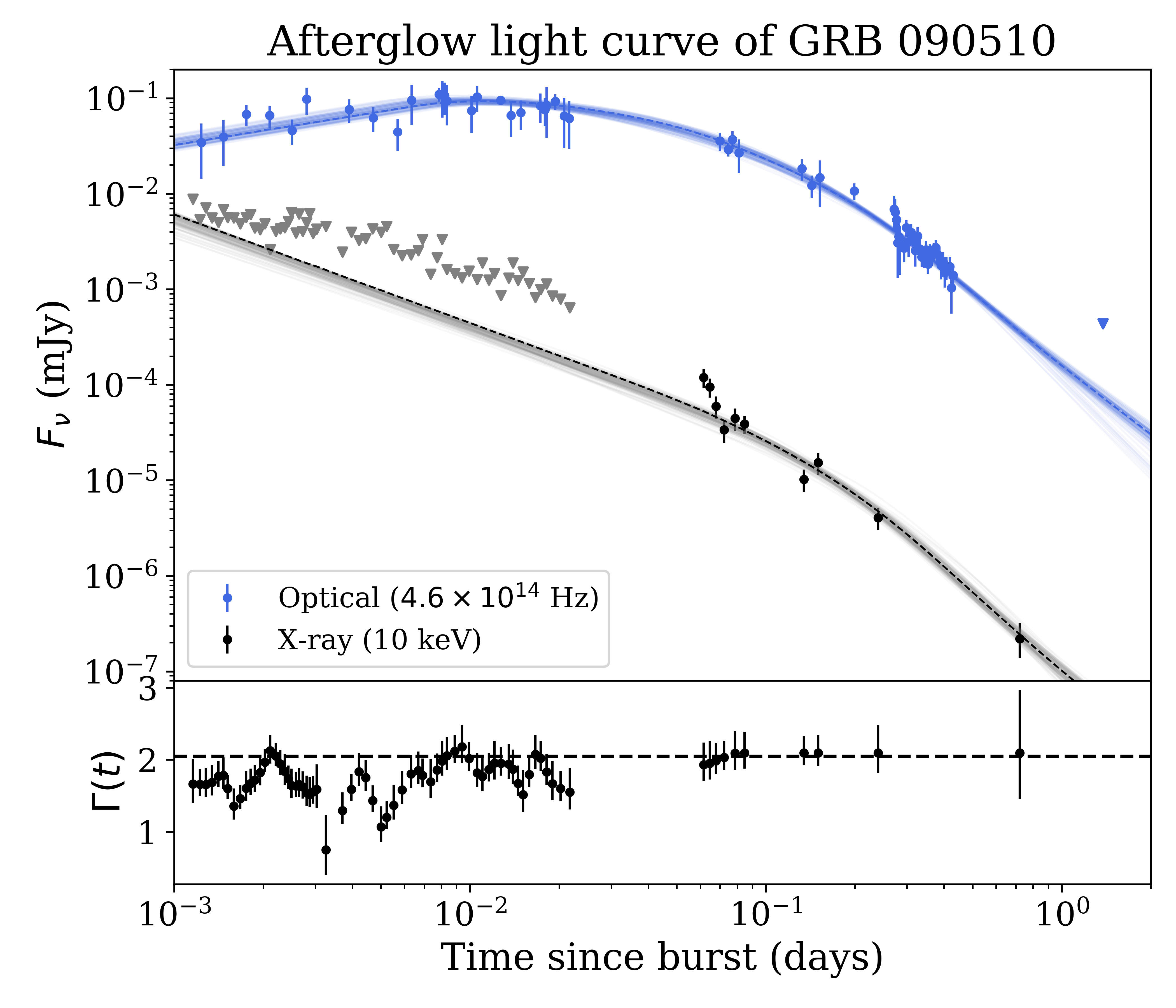}
    \includegraphics[width=\columnwidth]{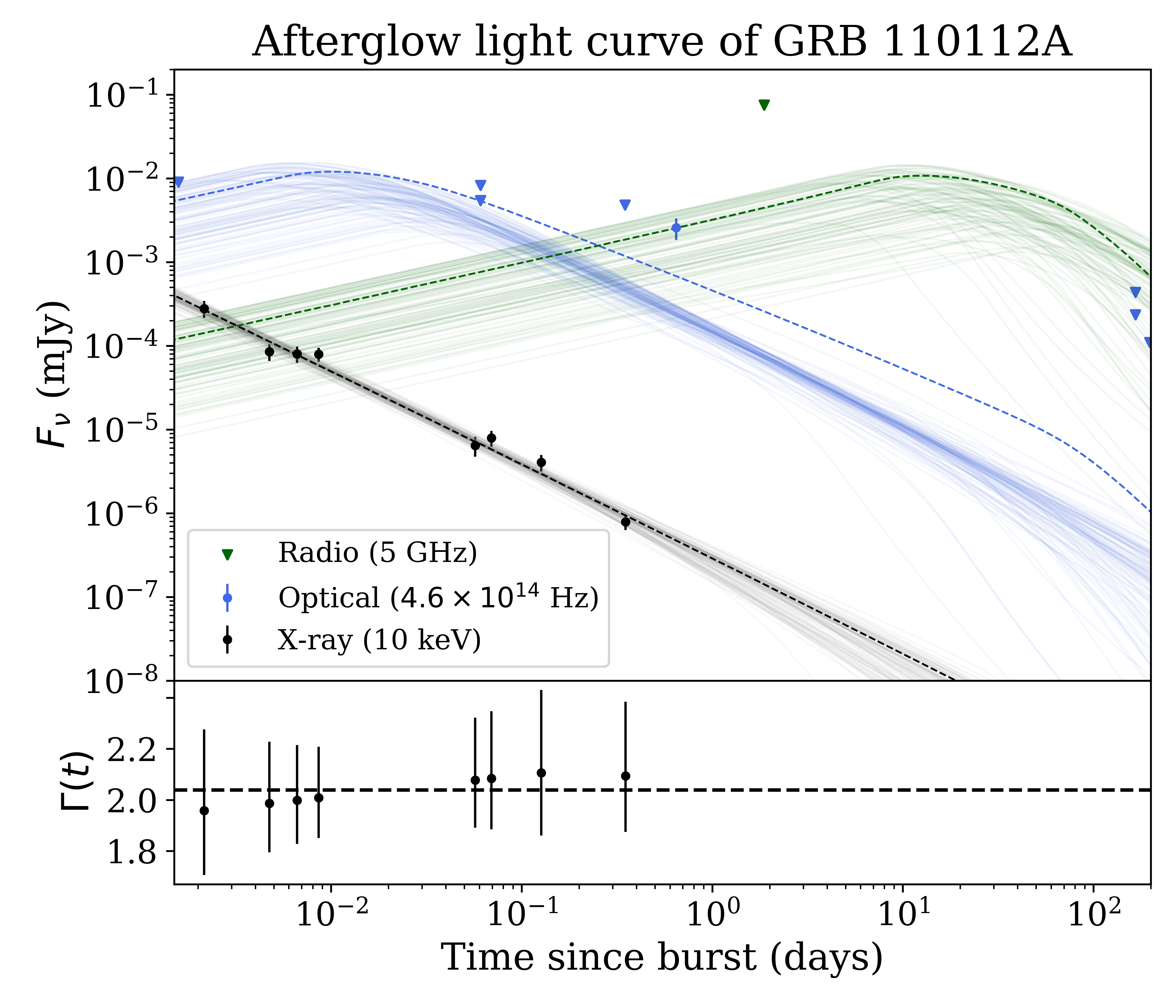}
\caption{continued.}
\end{figure*}
\begin{figure*}\ContinuedFloat
    \includegraphics[width=\columnwidth]{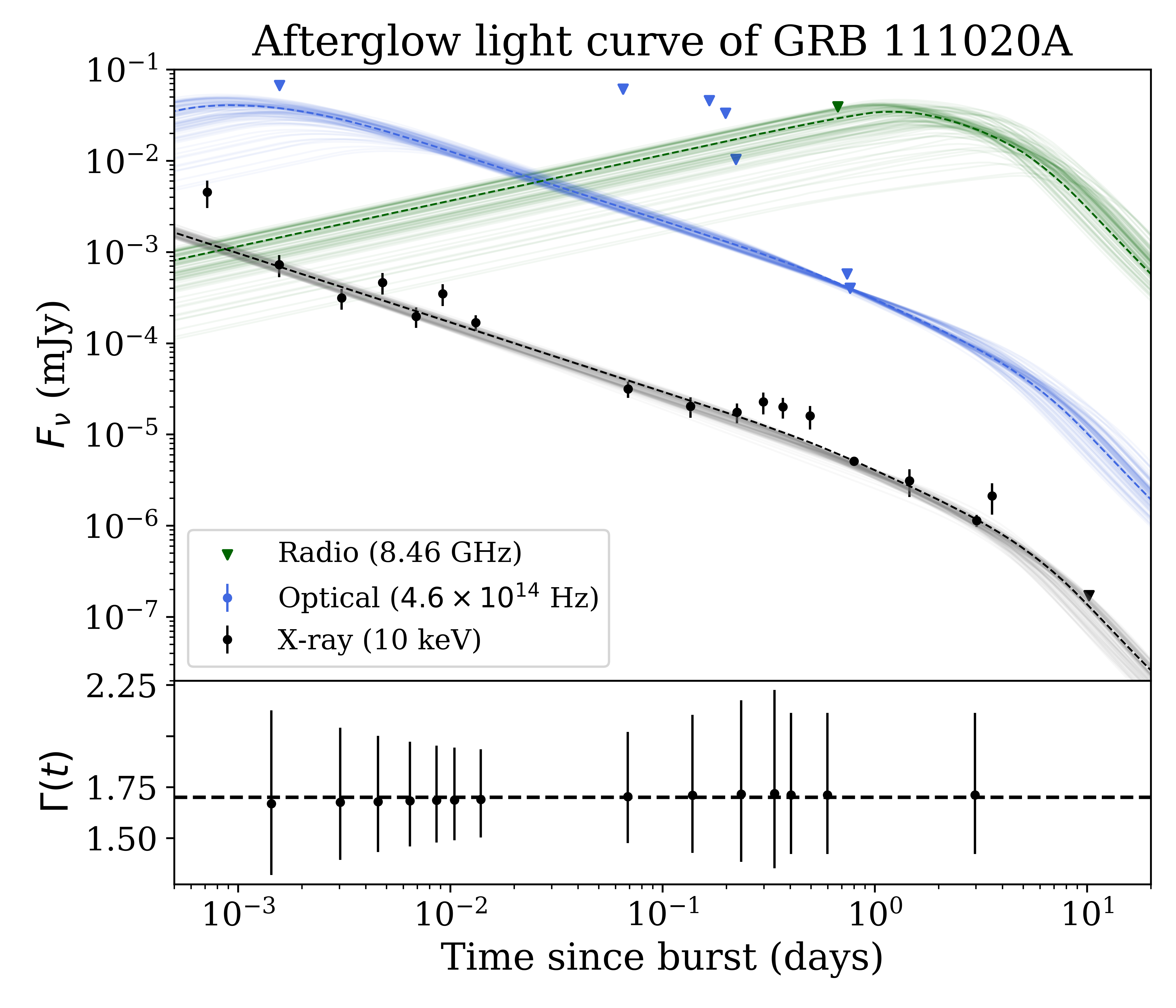}
    \includegraphics[width=\columnwidth]{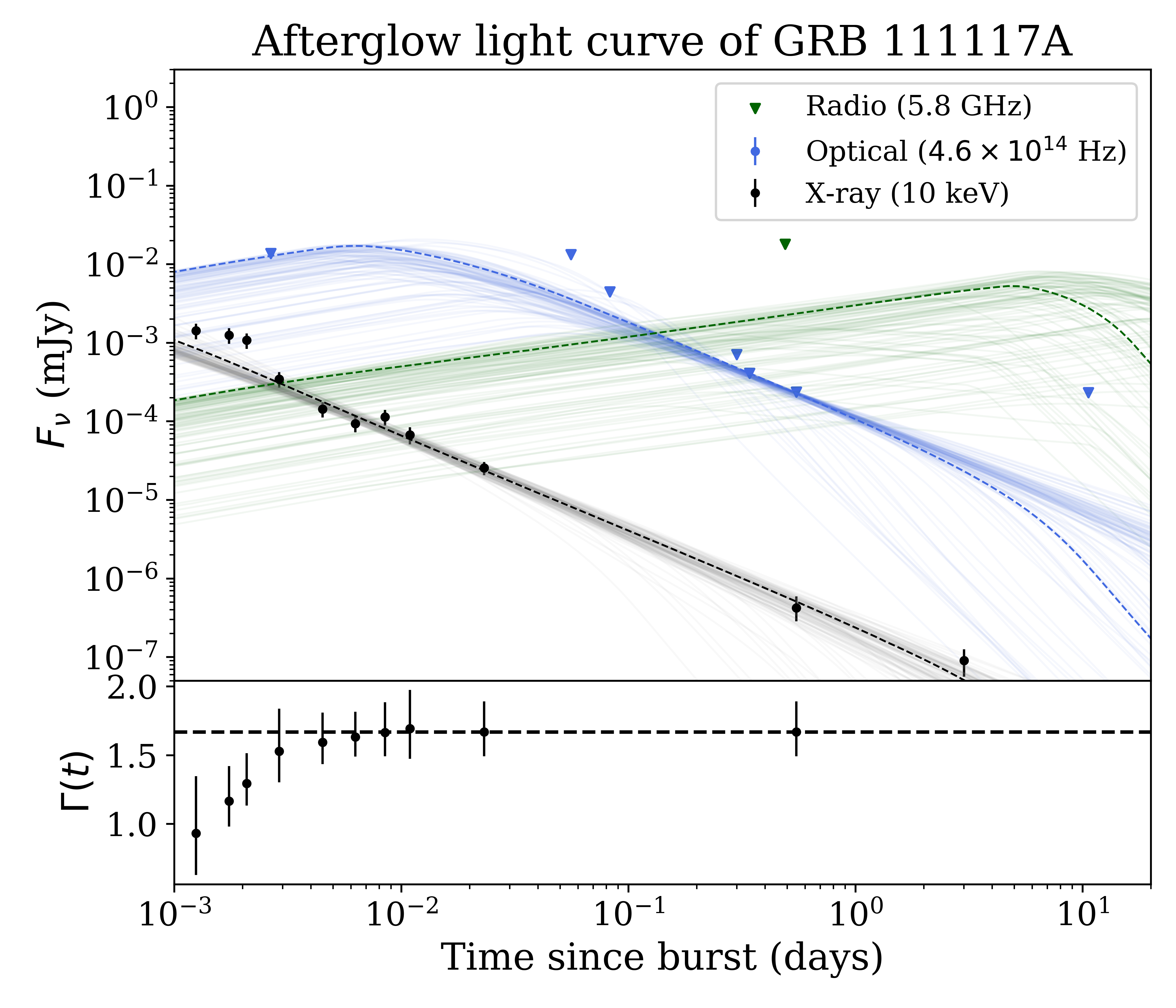}
    \includegraphics[width=\columnwidth]{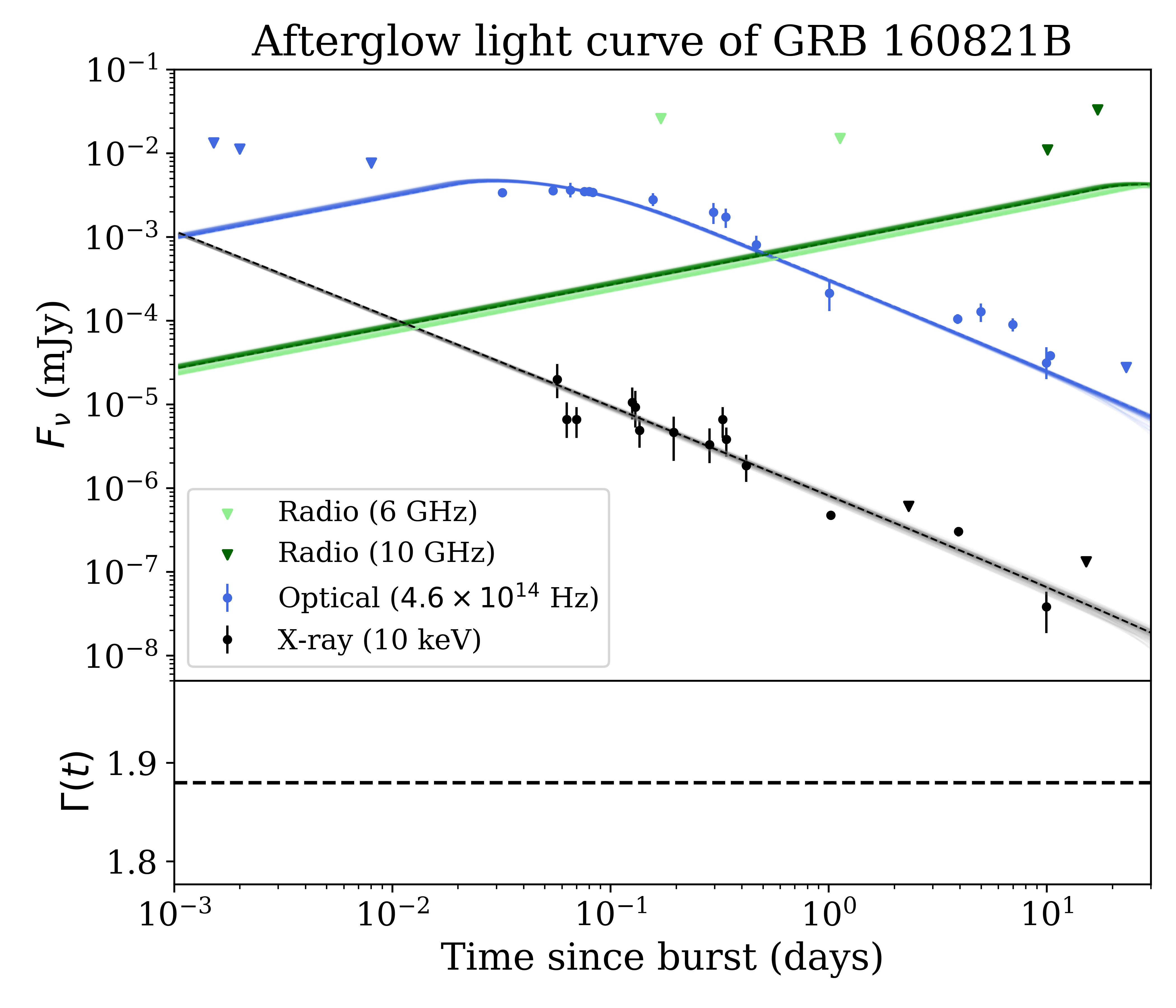}
    \caption{continued.}
\label{fig:fits}
\end{figure*}

\newpage
\begin{table*}
    \footnotesize
    \caption{Fitting results of broadband modelling of merger-induced GRB light curves with \texttt{afterglowpy} given a uniform jet model.}
    \makebox[1 \textwidth][c]{
    \resizebox{1.1 \textwidth}{!}{
    
    \begin{tabular}{ccccccccc} \hline
     GRB & $\theta_{\text{obs}}$ & log $E_{0}$ & $\theta_{\rm C}$ & log $n_{0}$ & $p$ & log $\epsilon_{e}$ & log $\epsilon_{B}$ & stat/d.o.f. (d.o.f.)   \\ 
     &(rad)  & (erg) & (rad) & (cm$^{-3}$) & & & & \\ \hline
     \vspace{0.1cm}
     050709 & $0.272_{-0.131}^{+0.142}$ & $50.75_{-0.15}^{+0.17}$ & $0.340_{-0.134}^{+0.126}$ & $-4.17_{-0.28}^{+0.31}$ & $2.95_{-0.07}^{+0.04}$ & $-0.52_{-0.05}^{+0.03}$ & $-0.60_{-0.15}^{+0.09}$ & 6.29 (3)   \\ 
     \vspace{0.1cm}
     050724 & $0.027_{-0.019}^{+0.036}$ & $50.34_{-0.13}^{+0.20}$ & $0.288_{-0.063}^{+0.050}$ &  $0.38_{-0.60}^{+0.39}$ & $2.29_{-0.09}^{+0.10}$ & $-0.61_{-0.17}^{+0.09}$ & $-1.23_{-0.46}^{+0.49}$ & 2.16 (8)  \\ 
     \vspace{0.1cm}
     051227 & $0.019_{-0.011}^{+0.013}$ & $52.86_{-0.67}^{+0.74}$ & $0.048_{-0.020}^{+0.021}$ & $-0.18_{-1.85}^{+0.88}$ & $2.16_{-0.09}^{+0.08}$ & $-1.04_{-0.69}^{+0.41}$ & $-4.52_{-0.93}^{+1.21}$ & 1.16 (3)  \\ 
     \vspace{0.1cm}
     061006 & $0.151_{-0.107}^{+0.159}$ & $53.02_{-0.10}^{+0.70}$ & $0.373_{-0.157}^{+0.108}$ & $-2.80_{-2.78}^{+2.40}$ & $2.14 \pm 0.06$ & $-1.42_{-0.94}^{+0.67}$ & $-4.04_{-1.83}^{+1.85}$ & 6.72 (4)  \\ 
     \vspace{0.1cm}
     061201 & $0.004_{-0.003}^{+0.006}$ & $50.34_{-0.06}^{+0.09}$ & $0.034\pm 0.004$ & $-4.52_{-0.22}^{+0.18}$ & $2.32_{-0.04}^{+0.05}$ & $-0.50_{-0.04}^{+0.02}$ & $-0.56_{-0.14}^{+0.06}$ & 3.77 (11)  \\ 
     \vspace{0.1cm}
     070714B & $0.003_{-0.002}^{+0.004}$ & $52.54_{-0.38}^{+0.66}$ & $0.018_{-0.008}^{+0.010}$ & $-2.67_{-1.29}^{+0.90}$ & $2.34_{-0.09}^{+0.08}$ & $-0.65_{-0.26}^{+0.13}$ & $-3.01_{-0.74}^{+0.97}$ & 1.88 (10)  \\ 
     \vspace{0.1cm}
     070724A & $0.008_{-0.006}^{+0.012}$ & $52.96_{-0.80}^{+0.69}$ & $0.066_{-0.027}^{+0.031}$ & $-0.22_{-1.87}^{+0.91}$ & $2.02_{-0.01}^{+0.02}$ & $-1.09_{-0.75}^{+0.44}$ & $-4.98_{-1.20}^{+1.31}$ & 3.97 (2) \\ 
     \vspace{0.1cm}
     071227 & $0.175_{-0.125}^{+0.167}$ & $50.78_{-0.47}^{+1.93}$ & $0.339_{-0.164}^{+0.130}$ & $-2.20_{-0.97}^{+1.92}$ & $2.08_{-0.06}^{+0.13}$ & $-0.87_{-1.57}^{+0.30}$ & $-1.38_{-1.52}^{+0.67}$ & 2.10 (2)  \\ 
     \vspace{0.1cm}
     090510 & $0.004_{-0.002}^{+0.003}$ & $52.10_{-0.17}^{+0.23}$ & $0.025_{-0.008}^{+0.010}$ & $-2.22_{-1.07}^{+0.92}$ & $2.16_{-0.06}^{+0.07}$ & $-0.65_{-0.23}^{+0.13}$ & $-1.70_{-0.70}^{+0.72}$ & 0.85 (68) \\ 
     \vspace{0.1cm}
     110112A & $0.152_{-0.109}^{+0.162}$ & $51.67_{-0.79}^{+1.17}$ & $0.371_{-0.160}^{+0.110}$ & $-4.49_{-1.03}^{+1.31}$ & $2.45_{-0.12}^{+0.07}$ & $-0.70_{-0.26}^{+0.16}$ & $-2.13_{-1.72}^{+1.16}$ &  7.31 (2) \\  
     \vspace{0.1cm}
     111020A & $0.015_{-0.007}^{+0.010}$ & $52.62_{-0.35}^{+0.53}$ & $0.036_{-0.011}^{+0.022}$ & $-4.70_{-0.89}^{+1.16}$ & $2.02\pm 0.01$ & $-0.59_{-0.16}^{+0.09}$ & $-2.28_{-0.90}^{+0.80}$ & 2.61 (10)  \\ 
     \vspace{0.1cm}
     111117A & $0.101_{-0.087}^{+0.258}$ & $52.49_{-0.41}^{+0.67}$ & $0.108_{-0.088}^{+0.258}$ & $-2.12_{-1.89}^{+1.39}$ & $2.39_{-0.12}^{+0.09}$ & $-0.64_{-0.26}^{+0.12}$ & $-3.40_{-1.08}^{+1.39}$ & 5.54 (4)  \\ 
     \vspace{0.1cm}
     160821B & $0.127_{-0.092}^{+0.132}$ & $50.46_{-0.03}^{+0.04}$ & $0.397_{-0.130}^{+0.090}$ & $-5.97_{-0.02}^{+0.05}$ & $2.39\pm 0.01$ & $-0.48_{-0.007}^{+0.003}$ & $-0.51_{-0.06}^{+0.03}$ & 4.21 (23)
    \end{tabular}
    }
    }
    \tablefoot{Reported are the median and 16th, 84th percentile values of the posterior distributions of the jet parameters of each GRB in the sample, as well as the fit statistic reduced by the number of degrees of freedom (d.o.f) corresponding to the \textit{maximum a posteriori} (MAP) parameter set.}
\label{tab:fits}
\end{table*}

\end{appendix}

\end{document}